\documentclass[a4paper,11pt]{amsart}

\DeclareRobustCommand{\SkipTocEntry}[5]{}

\usepackage{ amsmath }
\usepackage{marvosym}

\usepackage[T1]{fontenc}
\usepackage{empheq}
\usepackage{relsize}
\usepackage{bm}

\usepackage{upgreek}
\usepackage{ esint }
\usepackage{color}
\usepackage{comment}
\usepackage{amssymb}
\usepackage{amsfonts}
\usepackage{graphicx}

\usepackage{amscd}

\usepackage{slashed}
\usepackage{pdflscape}
\usepackage{tikz}
\usepackage{enumerate}
\usepackage{multirow}
\usepackage{stmaryrd}

\usepackage[linkcolor=blue]{hyperref}
\usepackage{mathrsfs}
\usepackage[colorinlistoftodos]{todonotes}

\usepackage[T1]{fontenc}

\definecolor{blue}{rgb}{.255,.41,.884} 
\definecolor{red}{rgb}{1, 0, 0} 
\definecolor{green}{rgb}{.196,.804,.196} 
\definecolor{yellow}{rgb}{1,.648,0} 
\definecolor{pink}{rgb}{1,0.5,0.5}

\newtheorem{theorem}{Theorem}[section]

\theoremstyle{definition}
\newtheorem{definition}[theorem]{Definition}
\newtheorem{example}[theorem]{Example}

\theoremstyle{remark}
\newtheorem{remark}[theorem]{Remark}

\usepackage{pdfsync}

\usepackage{graphicx} 
\usepackage{amssymb}

\newcommand{\be}{\begin{equation}}

\newcommand{\ee}{\end{equation}}

\newcommand{\C}{{\bf C }}

\newcommand{\ba}{\begin{array}}

\newcommand{\ea}{\end{array}}

\newcommand{\beq}{\begin{eqnarray}}

\newcommand{\eeq}{\end{eqnarray}}

\newtheorem{lm}{lemma}

\newtheorem{thee}{theorem}

\newtheorem{proo}{proposition}

\newtheorem{co}{corollary}

\newtheorem{rem}{remark}

\newtheorem{deff}{definition}

\newcommand{\bd}{\begin{deff}}

\newcommand{\ed}{\end{deff}}

\newcommand{\bl}{\begin{lm}}

\newcommand{\el}{\end{lm}}

\newcommand{\bp}{\begin{proo}}

\newcommand{\ep}{\end{proo}}

\newcommand{\bt}{\begin{thee}}

\newcommand{\et}{\end{thee}}

\newcommand{\bc}{\begin{co}}

\newcommand{\ec}{\end{co}}

\newcommand{\brm}{\begin{rem}}

\newcommand{\erm}{\end{rem}}

\newcommand{\f}{\overline{f}}

\usepackage{t1enc}

\newcommand{\newc}{\newcommand}

\renewcommand{\exp}{\operatorname{exp}}

\let\ccdot.

\newc{\aR}{\mbox{\boldmath{$ R$}}}
\newc{\aS}{\mbox{\boldmath{$ S$}}}
\newc{\aT}{\mbox{\boldmath{$ T$}}}
\newc{\aW}{\mbox{\boldmath{$ W$}}}

\newc{\aD}{\mbox{\boldmath{$ D$}}\hspace{-.2mm}}

\renewcommand{\colon}{\scalebox{1.2}{:}}

\newc{\aK}{\mbox{\boldmath{$ K$}}}
\newc{\aL}{\mbox{\boldmath{$ L$}}}

\usepackage{amssymb}
\usepackage{amscd}

\newcommand{\nn}[1]{(\ref{#1})}

\let\t=\tau

\newc{\obstrn}[2]{B^{#1}_{#2}}

\newcommand{\rpl}                         
{\mbox{$
\begin{picture}(12.7,8)(-.5,-1)
\put(0,0.2){$+$}
\put(4.2,2.8){\oval(8,8)[r]}
\end{picture}$}}

\newcommand{\lpl}                         
{\mbox{$
\begin{picture}(12.7,8)(-.5,-1)
\put(2,0.2){$+$}
\put(6.2,2.8){\oval(8,8)[l]}
\end{picture}$}}

\usepackage{ifthen}

\newc{\tensor}[1]{#1}
\newc{\Mvariable}[1]{\mbox{#1}}
\newc{\down}[1]{{}_{#1}}
\newc{\up}[1]{{}^{#1}}

\newc{\JulyStrut}{\rule{0mm}{6mm}}
\newc{\midtenPan}{\mbox{\sf S}}
\newc{\midten}{\mbox{\sf T}}
\newc{\midtenEi}{\mbox{\sf U}}
\newc{\ATen}{\mbox{\sf E}}
\newc{\BTen}{\mbox{\sf F}}
\newc{\CTen}{\mbox{\sf G}}

\def\sideremark#1{\ifvmode\leavevmode\fi\vadjust{\vbox to0pt{\vss
 \hbox to 0pt{\hskip\hsize\hskip1em
 \vbox{\hsize2cm\tiny\raggedright\pretolerance10000
  \noindent #1\hfill}\hss}\vbox to8pt{\vfil}\vss}}}

\numberwithin{equation}{section}

\newcommand{\hh}{{\hspace{.3mm}}}

\renewcommand\colon{\scalebox{1.3}{$:$}}

\renewcommand\geq{\geqslant}

 \newcommand{\bdot }{{\mathop{\lower0.33ex\hbox{\LARGE$\cdot$}}}}

\newcommand{\superimpose}[2]{%
  {\ooalign{$#1\@firstoftwo#2$\cr\hfil$#1\@secondoftwo#2$\hfil\cr}}}

\begin{document}

\renewcommand{\today}{}
\title{
{
 {The Quantum Darboux Theorem
 }}}
\author{ O. Corradini${}^\sharp$, 
E.~Latini${}^\flat$
\&  Andrew Waldron${}^\natural$}

\address{${}^\sharp$ 
Dipartimento di Scienze Fisiche, Informatiche e Matematiche,
Universit\`a degli Studi di Modena e Reggio Emilia, Via Campi 213/A, I-41125 Modena
\& INFN, Sezione di Bologna, Via Irnerio 46, I-40126 Bologna, Italy
 }
 \email{olindo.corradini@unimore.it}
 

\address{${}^\flat$ 
 Dipartimento di Matematica, Universit\`a di Bologna, Piazza di Porta S. Donato 5,
 and  INFN, Sezione di Bologna, Via Irnerio 46, I-40126 Bologna,  Italy}
 \email{emanuele.latini@UniBo.it}

  \address{${}^{\natural}$
  Center for Quantum Mathematics and Physics (QMAP)\\
  Department of Mathematics\\ 
  University of California\\
  Davis, CA95616, USA} \email{wally@math.ucdavis.edu}

\vspace{10pt}

\renewcommand{\arraystretch}{1}

%
%
%
%

%
%
%
%
%
%


\begin{abstract}

\noindent
The problem of computing quantum mechanical propagators can be recast as a computation of a Wilson line operator for parallel transport by a flat connection acting on a vector bundle of wavefunctions. In this picture the base manifold is an odd-dimensional symplectic geometry, or quite generically a contact manifold that can be viewed as a ``phase-spacetime'', while the fibers are Hilbert spaces. This approach enjoys  a ``quantum Darboux theorem'' that parallels the Darboux theorem on contact manifolds which turns local classical dynamics into straight lines.
We detail how the quantum Darboux theorem works for anharmonic quantum potentials. In particular, we develop a novel diagrammatic approach for computing the asymptotics of a gauge transformation that locally makes complicated quantum dynamics trivial. 

 \end{abstract}


\maketitle

\pagestyle{myheadings} \markboth{Corradini, Latini \& Waldron}{Quantum Darboux}


\tableofcontents

\section{Introduction}
A fundamental problem in quantum mechanics is to compute correlators
$$
\langle f| e^{-\frac i\hbar {\hat H (t_f-t_i)} } |i\rangle\, ,
$$
where the states $|i\rangle$, $|f\rangle$ are elements of some Hilbert space~${\mathcal H}$, the operator $\hat H$ is a quantum Hamiltonian, and $t_i$, $t_f$ are classical times measured in some laboratory. It is useful to view the coordinates $t_i$, $t_f$ as labels belonging to points in an odd-dimensional {\it phase-spacetime} manifold $Z$, that corresponds to all possible classical laboratory measurements of generalized times, positions and momenta. Then  the unitary time evolution operator
$\exp\big(\!-\frac i\hbar{\hat H (t_f-t_i)} \big)$
can be replaced by a Wilson line
$$
P_\gamma \exp\Big(-\int_\gamma \hat A\Big)\, .
$$
Here $\hat A$ is the connection form for a connection $\nabla=d+\hat A$
acting on sections of a Hilbert bundle (see~\cite{Dupre} and also~\cite{Krysl}) ${\mathcal H} Z$, over the base manifold  $Z$ with Hilbert space fibers ${\mathcal H}$.
Requiring that $\nabla$ is flat (and if necessary, quotienting by non-trivial holonomies), the correlator
$  \langle f|  P_\gamma \exp\Big(-\int_\gamma \hat A\Big) |i\rangle$ only depends on the endpoints of $\gamma$ and points $|i\rangle$, $|f\rangle$ in the fibers of ${\mathcal H}Z$ above these two  endpoints: 

\medskip

\vspace{-.2cm}
\begin{center}
\begin{picture}(100,100)(0,80)
\put(-10,35){
\includegraphics[width=5cm, height=5cm]
{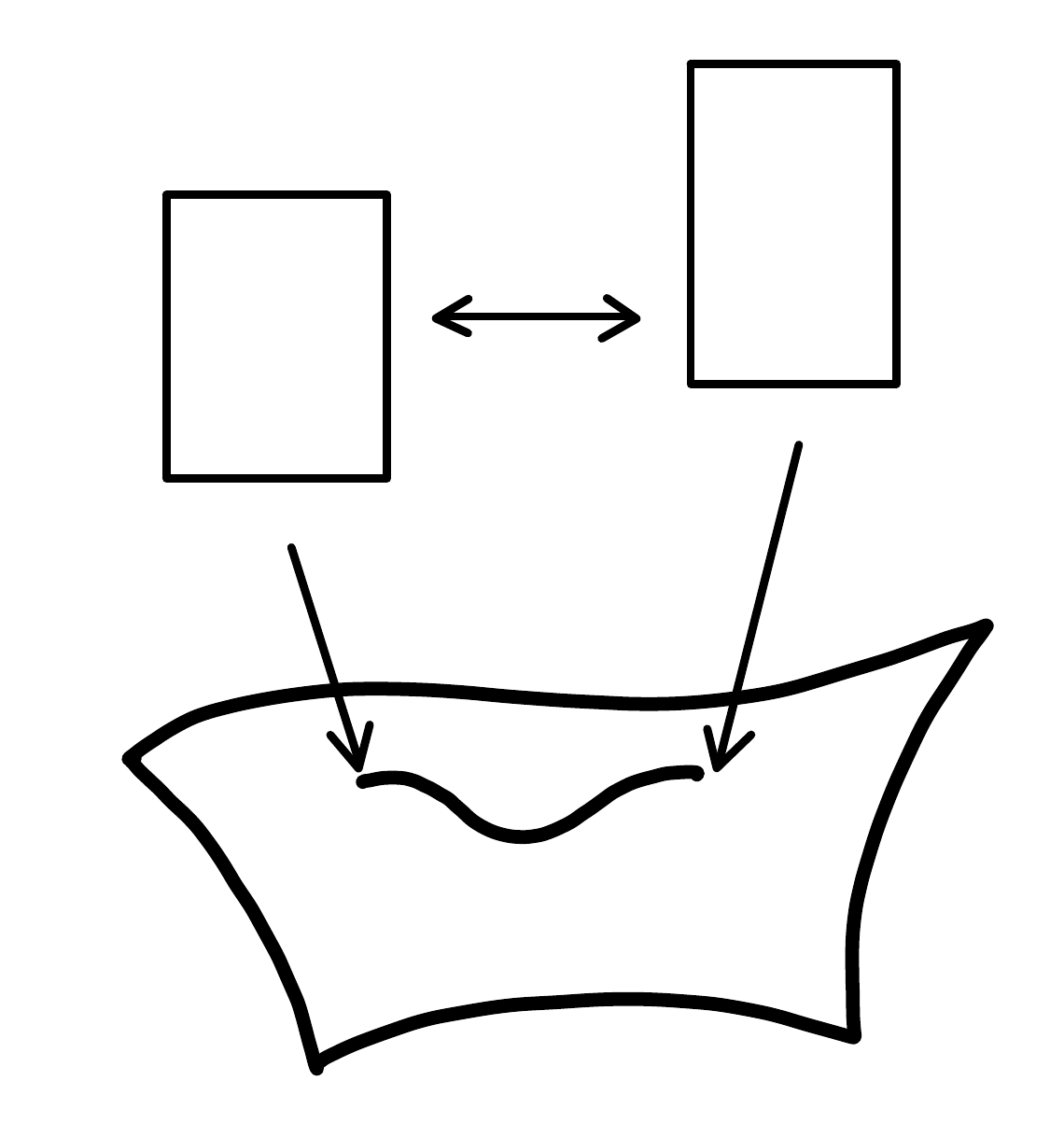}}
\put(-60,147){\scalebox{.9}{Hilbert space ${\mathcal H}$}}
\put(115,47){\scalebox{.9}{Phase-spacetime $Z$}}
\put(39,70){$\scriptstyle t_i$}
\put(87,71){$\scriptstyle t_f$}
\put(61,140){$\nabla$}
\put(27,133){$\scriptstyle |i\rangle$}
\put(95,149){$\scriptstyle |f\rangle$}
\put(62,64){$\gamma$}
\end{picture}
\end{center}
\vspace{1.2cm}
\bigskip

\noindent
We attach the moniker {\it quantum dynamical system} to the data $(\mathcal HZ,\nabla)$ of a Hilbert bundle equipped with a flat connection $\nabla$. The goal of this article is to use the gauge covariance of Wilson line operators, to map complicated quantum dynamics to simpler ones.

\medskip

The  article is structured as follows: We first review how to formulate a classical dynamical system in terms of an odd-dimensional analog of a symplectic geometry. We then explain, following~\cite{Her}, how to obtain a quantum dynamical system from the BRST quantization of a classical dynamical system. Here the BRST charge corresponds to the flat connection $\nabla$. We then treat Hamiltonian--Jacobi theory in terms of contact geometry.  (Note that a modern treatment of Hamilton--Jacobi actions  and their formal quantization is given in~\cite{Catt}.) Thereafter we describe the quantum analog of the Darboux theorem for quantum dynamical systems~\cite{Herla}.
After that we  primarily focus on quantum mechanics of a particle on a line. The underlying classical dynamical system gives evolution on a three-dimensional phase-spacetime manifold $Z$, for which we give an explicit account of the (classical) Darboux theorem. The remainder of the article is devoted to developing a diagrammatic calculus for computing the asymptotics of the gauge transformations appearing in the quantum Darboux theorem of~\cite{Herla}, and applying these to the computation of correlators.

\section{Dynamical Systems}

\noindent
Quite generally,  dynamics is a rule governing the time evolution of a system, so a dynamical system is a one-parameter family of maps $\Phi_\tau:Z\to Z$ where points $\tau\in {\mathbb R}$ are viewed as times and $Z$ is the state space of the system. Fixing a point $z_0\in Z$ and allowing~$\tau$ to vary gives a parameterized path $\gamma:{\mathbb R}\to Z$ in $Z$ with initial condition~$z_0$. Thus,   dynamics amounts to a set of parameterized paths (locally) foliating the space~$Z$. Often,~$Z$ is taken to be a symplectic manifold or phase space, in which case it is even-dimensional and points in that space label generalized positions and momenta. However, to maintain general covariance also with respect to choices of clocks, we shall demand that~$Z$ is an odd dimensional manifold which we term 
a {\it phase-spacetime}. We shall therefore view dynamics as a set of {\it unparameterized} paths (locally) foliating the odd-dimensional phase-spacetime~$Z^{2n+1}$.

\medskip

In general it is unusual to know explicitly the map $\Phi_{\tau}$ determining a dynamical system, instead one has some kind of local rule generating dynamics. For example, on a phase-space or symplectic manifold one typically considers a  Hamiltonian vector field~$X_H$ determined by a Hamiltonian function $H$ whose integral flow determines dynamics in terms of  {\it parameterized} paths. However, 
for $Z$ a phase-spacetime, we only need determine unparameterized geodesics. For that a local rule is generated by a maximally non-degenerate (so rank $2n$) two-form $\varphi \in \Omega^2 Z$, because an unparametrized  path $\gamma$ can be determined from the equation of motion 
\begin{equation}\label{eom}
\varphi(\dot \gamma,\cdot)=0\, .
\end{equation}
In the above $\dot \gamma$ is the tangent vector to $\gamma$ with respect to {\it any} choice of parameterization.

\begin{example} {\it Massless relativistic particle:}
Let $Z={\mathbb R}^7\ni (\vec k, x^0,\vec x)$ and 
$$
\varphi = d\vec k\,  \wedge\!\! \!\!\!\!\hh\cdot\, \,  (d\vec x + \hat k dx^0)\, .
$$
Then Equation~\nn{eom} is solved by
$$
\dot \gamma \propto
\frac{\partial}{\partial x^0} 
-\hat k \cdot \frac{\partial}{\partial \vec x}\, , 
$$
which is the tangent vector to the trajectory of a massless particle in Minkowski space moving in the direction $\hat k$.
\end{example}

\medskip

When the two-form $\varphi$ 
is closed we may write down an action principle for local dynamics, because locally $\varphi = d \alpha$ for some $\alpha \in \Omega^1 Z$. Then the action 
$$
S[\gamma] =\int_\gamma \alpha
$$
is extremized under compactly supported variations of the path $\gamma$ precisely when the Equation of Motion~\nn{eom} holds. Changing $\alpha$ by an exact term $d\beta$ does not change the equation of motion. Because we wish to focus on systems with an action principle, from now on, we shall assume that $\varphi$ is closed. 
When $\varphi\in \Omega^2Z$ is closed and maximally non-degenerate, we call $(Z,\varphi)$ a {\it dynamical phase-spacetime}. 
The closed, maximally non-degenerate two form $\varphi$ is termed an {\it odd symplectic form}.
(Sometimes the additional data of a volume form is added to the definition of an odd symplectic manifold, see~\cite{He}.)
The data of $\alpha$ modulo gauge transformations $\alpha\mapsto\alpha + d\beta$, whose curvature $\varphi=d\alpha$ is maximally non-degenerate,  also determines the dynamics, and we shall call this a {\it dynamical connection}.

\medskip
Given a dynamical system $(Z,\varphi)$, a function $Q\in C^\infty Z$ is said to be {\it conserved} when 
$$
{\mathcal L}_\rho Q = 0\, ,
$$
for any vector field $\rho$ obeying $\varphi(\rho,\cdot)=0$. Note that 
${\mathcal L}_{f\rho} Q=f{\mathcal L}_{\rho} Q$ for any $f\in C^\infty Z$. A vector field $u$ is said to generate a {\it symmetry} when 
$$
{\mathcal L}_u \varphi = 0\, .
$$
Conserved quantities and symmetries obey a Noether theorem: Given $Q$ conserved, any solution $u$ to 
\begin{equation}\label{clQ}
dQ=\varphi(u,\cdot)
\end{equation}
is a symmetry because ${\mathcal L}_u \varphi = d\,  \iota_u \varphi =d^2 Q= 0$. Vector fields $f\rho$ correspond to trivial (constant) conserved charges. The equation displayed above is always solvable for $u$ because $\iota_\rho dQ=0$, and the solution for $u$ is unique modulo a term $f\rho$. Conversely, given a symmetry $u$, then the one-form $\varphi(u,\cdot)$ is exact because
$d\iota_u \varphi = {\mathcal L}_u\varphi = 0$. Hence, locally, there always exists a smooth conserved $Q$ solving the above display.

\medskip

Experimentally, the initial conditions for a dynamical system cannot be determined with infinite precision. Instead, one associates an experimental uncertainty to an
open ball in  $Z$ about about some initial point $z_0\in Z$.
Therefore, a notion of volume is needed to ascertain the accuracy of such an initial measurement.
Because $\varphi$ is maximally non-degenerate, the form $\varphi^{\wedge n}\in \Omega^{2n}Z$ is non-vanishing. To make a top form, we need to exterior multiply this by some one-form. When $\varphi = d\alpha$---which is certainly true locally and for many physical systems and most experimental  apparatus can anyway only handle a measurements in a finite range---a natural choice is the one-form~$\alpha$, in which case we define
$$
\operatorname{Vol}_\alpha = \alpha\wedge \varphi^{\wedge n}\in \Omega^{2n+1} Z\, .
$$
{\it A priori}, the ``volume form'' $\operatorname{Vol}_\alpha$ need not be non-degenerate, and moreover it depends, up to an exact term, on the choice of $\alpha$; $\operatorname{Vol}_{\alpha+d\beta}=\operatorname{Vol}_\alpha+d(\beta\varphi^{\wedge n})$.
For some systems, it may suffice only to have a non-degenerate volume form defined on suitable hypersurfaces of $Z$ by $\varphi^{\wedge n}$, but generally we 
focus on systems where $\operatorname{Vol}_\alpha$ is a volume form and view the choice of $\alpha$ as part of the data of the system required to give a good measurement theory.
The case where $\varphi$ is globally the $d$ of a given $\alpha$ and $\operatorname{Vol}_\alpha$ is non-degenerate is in some sense optimal. In this case $\alpha$ is called a {\it contact one-form}.
The data $(Z,\alpha)$ is a {\it strict contact structure} and the dynamics on $Z$
determined by the uniquely defined vector field  $\rho\in \Gamma(TZ)$ such that
$$
\varphi(\rho,\cdot)=0\, ,\qquad\alpha(\rho)=1\, ,
$$
are called {\it Reeb dynamics} and $\rho$ is the {\it Reeb vector field}. The canonical parameterization of path $\gamma$ in $Z$ such that 
$$
\dot \gamma = \rho
$$
is akin to that of geodesics in a Riemannian manifold by their proper length.
A {\it contact structure} $(Z,\xi)$ is the data of a hyperplane distribution $\xi$ in $TZ$ determined by the kernel of a contact one-form. A useful starting reference describing contact geometry is~\cite{Geiges}.

\begin{example}
{\it Massive relativistic particle:}
 Let $Z={\mathbb R}^3\ni(p,t,x)$ and 
 $$
 \alpha = pdx - \sqrt{p^2+m^2} \, dt \, .
 $$
 Then 
 $$
 \operatorname{Vol}_\alpha = \frac{m^2}{\sqrt{p^2+m^2}} \, dp \wedge dt\wedge dx\neq 0\, ,
 $$
 so $\alpha$ is a contact one-form. 
 It not difficult to verify that the Reeb trajectories correspond to those of a mass $m$, relativistic particle in two dimensional Minkowski space.
\end{example}


\section{Quantization}

\noindent
Our main focus is quantization of dynamical phase-spacetimes. We follow the treatment in~\cite{Her}.
The framework is quite close to that developed by  Fedosov for deformation quantization of symplectic manifolds~\cite{Fed} (see also~\cite{Grig}).
To quantize a dynamical phase-spacetime~$(Z,\varphi)$ we begin with an action on paths~$\gamma$ in $Z$
$$
S[\gamma]=\int_\gamma \alpha\, ,
$$
where $\alpha$ is any primitive of $\varphi$,
$$
\varphi = d\alpha\, .
$$
For locally supported variations of $\gamma$, the potential failure of $\alpha$ to be unique or to exist globally, is irrelevant. The above action can be reformulated as a Hamiltonian system on a coisotropic submanifold ${\mathcal C}$ of a larger symplectic manifold ${\mathcal Z}$, where
$$
{\mathcal Z}=T^*Z\oplus \xi^*\, .
$$
In the above, $\xi^*$ is the bundle of hyperplanes in $T^*Z$ defined by the kernel of the ray  defined by the kernel of $\varphi$. So, for example, if $\varphi(r,\cdot)=0$ for the ray $r$, then $\xi^*=\ker r$.
 In the case that $\alpha$ is a global contact form, this is the dual of the maximally non-integrable distribution $\xi$ defining a contact structure. Motivated by this correspondence, we call~$\xi^*$ a {\it codistribution}.
  The direct sum above is the  Whitney vector bundle sum, so ${\mathcal Z}$ has base $Z\ni z^i$ and fiber ${\mathbb R}^{2n+1}\oplus {\mathbb R}^{2n}\ni(p_i,s_a)$. 
We call ${\mathcal Z}$ an {\it extended phasespace}, it has a symplectic current (or Liouville form) $\Lambda$ obtained from the sum of the tautological one-form on $T^*Z$ and the standard Liouville form $\lambda_s$ on the ${\mathbb R}^{2n}$ fibers of~$\xi^*$.
In local coordinates $(z^i,p_i,s_a)$ for ${\mathcal Z}$, 
$$
\Lambda = p_i dz^i + \frac12 s_a j^{ab} ds_b\, .
$$
This makes ${\mathcal Z}$ into a symplectic manifold. 
Here $j^{ab}=-j^{ba}$ is an odd bilinear form defining the invariant tensor of $Sp(2n)$. Also, we employ the convention $X_a=j_{ab} X^b$. 
A coisotropic submanifold ${\mathcal C}$ of this, is determined by the $2n+1$ first class constraints
$$
C_i=p_i-A_i(z,s)\, ,
$$
where the one-form $A=\alpha + \kappa(z,s)+a(z,s)$ obeys the Cartan Maurer equation
$$
dA + \{A\stackrel\wedge, A\}_{PB}=0\, .
$$
Also, the leading fiberwise jet  of the one-form $\kappa(z,s)$ must obey the maximal rank condition
$$
\kappa(z,s) = e^a(z) s_a \, ,\qquad
\tfrac 12\,  j_{ab} e^a\wedge e^b = \varphi\, , 
$$
and
\begin{equation}\label{s2}
a(z,s)={\mathcal O}(s^2)\, . 
\end{equation}
In the above, the $2n$ one-forms $e^a$ are a basis, or adapted coframe,  for the codistribution~$\xi^*$, such that $e^a(\rho)=0$.
We shall often refer to these as {\it soldering forms}.
The extended action for paths $\Gamma$ in ${\mathcal \xi}$, obtained by integrating out the momenta $p_i$ by solving the constraints $C_i$, and given by
$$
S[\Gamma]=\int_\Gamma (\lambda_s +A)\, ,
$$
then  gives a (gauge) equivalent 
description of the dynamics given by the original action~$S[\gamma]$ above. Moreover, the constraints $C_i$ are abelian, so the classical BFV-BRST charge is simply $Q_{BRST}=c^i (p_i - A_i(s,z))$ where $c^i$ are ghosts. This charge is ``nilpotent'' (strictly its BFV Poisson bracket with itself vanishes) by virtue of the Cartan--Maurer equation.  In the above discussion we have effectively converted a system with a mixture of first and second class constraints to one with only first class constraints, this is an example of a general technique due to~\cite{BFVsecond}.

The extended action  easily quantized by viewing the ghosts $c^i$ as one-forms on the base manifold $Z$ (see for example~\cite{Witten}), so that~$\frac i \hbar \, c^ip_i$ becomes the exterior derivative $d$ acting on forms on~$Z$ and the fiber coordinates~$s_a$ become operators $\hat s_a\in \operatorname{End}{\mathcal H}$ 
acting on some choice of  Hilbert space ${\mathcal H}$ and subject to 
$$
[\hat s_a,\hat s_b]=i\hbar j_{ba}\, .
$$
In fact, since states are defined by complex rays,  we only need to consider the projective Hilbert space ${\mathbb P}({\mathcal H})\ni [|\psi\rangle]=[
[e^{i\theta}|\psi\rangle]$ (for any real $\theta$).
We will abbreviate the notation  ${\mathbb P}({\mathcal H})$  by
${\mathcal H}$ in what follows.

The BRST Hilbert space ${\mathcal H}_{BRST}$ is then 
differential forms on $Z$ taking values in ${\mathcal H}$. At ghost number zero, these
are sections of a Hilbert bundle ${\mathcal H}Z$ which is a vector bundle with fibers given by ${\mathcal H}$ and the base manifold is $Z$. In general BRST wavefunctions~$\Psi_{BRST}$ obey
$$
\Psi_{BRST}\in \Gamma({\mathcal  H}Z)\otimes\Omega Z={\mathcal H}_{BRST}\, .
$$
The ghost number grading for  space is given by form degree.

Note that the  frame bundle of the dual $\xi:=(\xi^*)^*$ of the codistribution $\xi^*$ canonically defines a principal $Sp(2n)$
bundle over $Z$, because $\varphi$ 
endows  $\xi$ with non-degenerate,
skew symmetric bilinear form. 
This canonically and globally defines 
the associated vector bundle 
with fibers given by ${\mathcal H}$.
In turn the principal bundle of orthonormal frames with respect to the Hilbert space inner product has structure group $U({\mathcal H})$. Let us denote by ${\mathcal H}Z$ the associated vector bundle with fibers ${\mathcal H}$ transforming under the fundamental representation of $U({\mathcal H})$. This is the bundle of wavefunctions defined up to unitary equivalence where the base manifold $Z$ plays the {\it r\^ole} of a generalized time coordinate.
.

\medskip 
The quantum BRST charge is a flat connection form on ${\mathcal H}Z$;  this extends by linearity to higher forms in  ${\mathcal H}_{BRST}$.
It is given by
$$
\frac i \hbar
\widehat Q_{BRST}= d +\hat A=:\nabla\, ,
$$
where  $\hat A$ is a one-form taking values in Hermitean operators on ${\mathcal H}$  is the quantization of the one-form $A$, 
and
similarly to the classical solution $A(z,s)$ to the Cartan--Maurer,  decomposes as
\begin{equation}\label{decompose}
\hat A = \frac{\alpha  \operatorname{Id}}{i\hbar} +\frac{\hat \kappa}{i\hbar}+  \hat a\, .
\end{equation}
We require that $\hat A \in \operatorname{End}(\Gamma({\mathcal H}Z))\otimes \Omega^1Z$ gives a solution of  the flatness condition
$$
\nabla^2=0\, .
$$
Given a connection form $d + \hat A$, the map
$$
\hat \kappa :\Gamma({\mathcal H}Z)\otimes \Gamma(TZ) \rightarrow 
\Gamma({\mathcal H}Z)\, ,
$$
is called the {\it quantum calibration map}. We require that it obeys the Heisenberg algebra, in the sense that for any $u,v\in  \Gamma(TZ)$
$$
\hat \kappa(u)\circ \hat \kappa(v) -
\hat \kappa(v)\circ \hat \kappa(u)=i\hbar\,  \varphi(u,v)\, .
$$
This can be solved locally using the coframe by writing $\hat \kappa = e^a \hat s_a$. 
Therefore this map calibrates quantum operators $\hat s$ to the underlying classical phase-spacetime manifold $Z$. Indeed the above map always exists, because ${\mathcal H}$ comes equipped with a representation of the Heisenberg algebra; this is  uniquely defined up unitary equivalence by the Stone von Neumann theorem. The map $\hat \kappa$ is a symplectic analog of Clifford multiplication for spinor bundles, see~\cite{Kostant}.

For any quantum system with a classical limit, or that arises via quantization, there must be a flow with respect to some parameter $\hbar$ encoding
how quantum quantities respond to changes in $\hbar$. Therefore, we introduce a grading operator
$$
{\sf gr}= 2 \hbar \frac{\partial}{\partial \hbar} + E\, ,
$$
where the operator $E:\Gamma({\mathcal H}Z)\to\Gamma({\mathcal H}Z)$ obeys
$$
E\circ \hat \kappa(u)-\hat \kappa(u) \circ  E = \hat \kappa(u)\, ,\quad \forall u\in \Gamma(TZ)\, .
$$
This grading operator is also needed to state the quantum analog of the classical higher jet condition in Equation~\nn{s2}.
For example, in any choice of polarization $\hat s_a=(\hat s^A, \hat s_B)$ such that
$[\hat s^A,\hat s_B]=i\hbar \, \delta^A_B$ acting on ${\mathcal H}=L^2({\mathbb R}^n)\ni \psi(s^A)$ with $\hat s^A \psi = s^A \psi$, $\hat s_A \psi = \frac\hbar i \partial \psi/\partial s^A$,
the map $E$ is given by 
$$
E=\frac{i}\hbar\sum_{A=1}^n\hat s^A \hat s_A+2\hbar \frac{\partial }{\partial \hbar}\, .
$$
Let us say that an operator $\hat O\in \operatorname{End}(\Gamma({\mathcal H}Z)$ has grade $k$ when
$$
{\sf gr}\circ \hat O + \hat O \circ {\sf gr} = k\, \hat O\, .
$$
We require that the operator $\hat a$ in Equation~\nn{decompose} has grade greater than or equal to zero.
Then, given the data of a dynamical connection $\alpha$ and a quantization map $\hat \kappa$, a solution 
$$
\nabla= \frac{\alpha  \operatorname{Id}}{i\hbar} +\frac{\hat \kappa}{i\hbar}+ d+ \hat a
$$
to the flatness condition $\nabla^2=0$ will be called a {\it quantum connection}. When  only a formal power series solution for $\nabla$ 
is given, we call this a {\it formal quantum connection}. 
Not that the first and third term of the above connection are the starting point for the geometric quantization of contact manifolds developed in~\cite{Fitz}.
Local existence of formal quantum connections 
given the data $(Z,\alpha, \hat \kappa, {\sf gr})$ is not difficult to establish~\cite{Her} (alternately, see the Darboux construction of a flat quantum connection given in Section~\ref{QDsec}). Indeed, 
when the calibration map is  given by coframes, 
 the operator $\hat a$ can be expanded order by order in the grading
$$
\hat a = \frac{1}{i\hbar } \!\!\sum_{\begin{array}{c}
\scriptstyle j, \ell\geq 0\\[-1mm]
\scriptstyle
j+2\ell\geq 2
\end{array}
}\!
\frac{ \hbar^\ell}{j!}
 \,  \omega^{a_1\ldots a_j}\hat s_{a_1}\cdots \hat s_{a_j}\, .
$$
In the above, the one-forms $\omega^{a_1\ldots a_j}$ are totally symmetric in the labels $a_1,\ldots, a_j$, and are determined by solving a system of algebraic, zero curvature,  equations.

\subsection{Quantum Darboux Theorem}\label{QDsec}

Locally, the contact Darboux theorem states that there exist coordinates $(\vec \pi, \vec  \chi, \psi)$ such that any contact form $\alpha$ can be written as
$$
\alpha=\vec \pi\cdot d\vec \chi - d\psi\, . 
$$
Here the Reeb vector is $\rho = -\frac\partial {\partial \psi}$ so that evolution is along straight lines
of constant $\vec \pi$ and $\vec \chi$. In the case where only a dynamical phase-spacetime $(Z,\varphi)$ is given, because $\varphi$ is closed we may always locally write $\varphi = d\alpha'$. Moreover, in the case that $\alpha'$ is not contact, because $\varphi$ is non degenerate, we may add to $\alpha'$ an exact term $d\beta$ such that $\alpha=\alpha'+d\beta$ is a
contact form, at least locally. 
Therefore, the contact Darboux theorem applies to dynamical phase-spacetimes as well.

There is a quantum analog of the 
Darboux theorem for formal quantum connections~\cite{Herla}. Before discussing this we need to talk about gauge transformations for quantum connections. In the previous section we stipulated that the quantum connection $\nabla$ was a flat connection form on the bundle ${\mathcal H}Z$. This is an associated vector bundle to a principal~$U({\mathcal H})$ bundle over $Z$. 
The connection form obeys a self-adjoint condition
$$
\hat A_u=\hat A_u^\dagger\in \operatorname{End}(\Gamma({\mathcal H}Z))\, ,
$$
for any $u\in \Gamma(TZ)$, and the adjoint is defined fiberwise using the adjoint of ${\mathcal H}$. 

The quantum Darboux theorem states that locally any pair of formal quantum connections for a given choice of dynamical phase-spacetime $(Z,\varphi)$ are 
formally gauge equivalent~\cite{Herla}. The result is established inductively in the grading ${\sf gr}$ by showing that there exists a formal  $U({\mathcal H})$ gauge transformation $\hat U$
such that 
\begin{equation}\label{QU}
\hat U \hh\nabla \hh\hat U^{-1} = \nabla_{\rm D}\, .
\end{equation}
Here $\nabla_{\rm D}$ is a quantum connection whose quantum calibration map $\hat \kappa$ is closed so that
$$
\nabla_{\rm D}
:=\frac{\alpha}{i\hbar} +
\frac{\hat \kappa}{i\hbar} +d
$$
is obviously flat.
To see that such a connection always exists locally, one can
use a Darboux coordinate ball, for which $\varphi = d\alpha$ and  where $\alpha$ is contact with enjoys coordinates $(\vec \pi, \vec \chi, \psi)$ such that
$$
\alpha = \vec \pi \cdot d\vec \chi - d\psi\, .
$$
Then the quantum calibration map $\hat \kappa = e^a \hat s_a$  can be built from closed coframes
$$
e^a = (d\vec \pi,d \vec \chi)\, .
$$
The quantum calibration map for a general quantum connection $\nabla$ will not be given in terms of closed frames, but an $Sp(2n)$
gauge transformation in the frame bundle of $\xi^*$ can be employed to achieve this. Locally, as discussed earlier, this lifts to an $Mp(2n)$ gauge transformation $\hat U_0$.
The strategy to find $\hat U$ is first to find  the metaplectic transformation~$\hat U_0$.
Then, 
\begin{equation}\label{a}
\nabla_0:=
\hat U_0^{-1} \nabla_{\rm D} \hat U_0
=\nabla + \hat a_0\, ,
\end{equation}
for some  $\hat a_{0}$ of grade $0$ or greater. Thereafter one solves for a form gauge transformation~$\hat U_1$ of grade one higher, given as a formal series in the grading, such that $\hat U_1^{-1} \nabla_0\hat U_1 = \nabla$. This means we must solve the equation
\begin{equation}\label{Schrod}
\nabla\hat U_1^{-1}=\hat U_1^{-1} \hh \hat a_0\, ,
\end{equation}
where $\nabla \hat U_1^{-1}:=[\nabla,\hat U_1^{-1}]$ is the adjoint action of  $\nabla$. Note that acting with $\nabla$ again on the above, just returns $\nabla_0^2=0$ 
as an integrability condition.
The above display has a formal (and possibly only local) solution for $\hat U_1^{-1}$~\cite{Herla}. 
The proof is by induction in the grading.


\medskip
A main aim of this article, is to study explicit, global (but possibly formal),  solutions to 
the above equation for quantum mechanical systems describing dynamics on a line. Knowledge of the gauge transformation $\hat U_1$ is powerful, because it relates non-trivial interacting systems to their trivial Darboux counterparts. The first step is to study the classical Darboux theorem for these models.

\section{Contact Hamilton--Jacobi Theory}

Let us consider a one-dimensional system with time-dependent Hamiltonian $H(p,q,t)$. 
The standard Hamilton--Jacobi theory for this system (see~\cite{Catt} for a modern treatment) can be recovered by studying diffeomorphisms on 
a three-dimensional dynamical phase-spacetime manifold $Z$ with local coordinates $(p,q,t)$ 
and odd symplectic form 
\begin{eqnarray*}
\varphi &=& \;\;\,  dp\wedge dq - dH\wedge dt\\[1mm]
&=&
\Big(dp+\frac{\partial H}{\partial q}\, dt\Big)\wedge
\Big(dq-\frac{\partial H}{\partial p}\, dt\Big)\, .
\end{eqnarray*}
This gives dynamics
$$
\dot \gamma\propto\frac{\partial }{\partial t }- \frac{\partial H}{\partial q }\frac{\partial }{\partial p }
+\frac{\partial H}{\partial p }\frac{\partial }{\partial q }\, ,
$$
where where $\propto$ denotes equality up to multiplication by some non-vanishing
function on~$Z$.
Calling the worldline parameter $\tau$, and choosing this function to be unity, the equations of motion are
$$
\frac{\partial t}{\partial \tau}=1\, ,\qquad
\frac{\partial p}{\partial \tau}=-\frac{\partial H}{\partial q }\, ,\qquad
\frac{\partial q}{\partial \tau}=\frac{\partial H}{\partial p }\, .
$$
In the  gauge $t(\tau)=\tau+c$ these are the standard Hamilton's equations.

\medskip

The odd symplectic form $\varphi$ can be written as the exterior derivative of the one-form
\begin{equation}\label{alphaH}
\alpha = pdq - H(p,q,t) dt\, .
\end{equation}
Away from the zero locus of $p\hh\frac{\partial H}{\partial p}-H$, the above form is in fact contact. 
The contact Darboux theorem ensures that locally we can find a new local coordinate system $(\uppi, \upchi, \uppsi)$ such that 
\begin{equation}\label{alphaD}
\alpha = \upchi d \uppi - d \uppsi\, .
\end{equation}
Our aim in this section is to give an (as) explicit (as possible) formula for the diffeomorphism bringing $\alpha $, at least  on some open set $U$, to its Darboux form displayed above.

\medskip

The Reeb vector  for $\alpha$ as in Equation~\nn{alphaH}, is given in these Darboux coordinates by~$\rho = -\frac{\partial}{\partial \uppsi}$, while in the original coordinate system
\begin{equation}\label{yourboat}
\Big(p\hh\frac{\partial H}{\partial p}-H\Big)\, 
\rho = \frac{\partial }{\partial t }- \frac{\partial H}{\partial q }\frac{\partial }{\partial p }
+\frac{\partial H}{\partial p }\frac{\partial }{\partial q }\, .
\end{equation}

\noindent
To proceed we need to know one conserved quantity $K\in C^\infty U$, 
\begin{equation}\label{conserved}
{\mathcal L}_\rho K =\iota_\rho dK=0\, .
\end{equation}
 This condition is always in principle {\it locally} solvable, but not in terms of explicit first integrals.
Explicitly, it
amounts to solving
$$
\frac{\partial K}{\partial t}=\{H,K\}_{_{\!\rm PB}}\, ,
$$
where $\{\cdot,\cdot\}_{_{\!\rm PB}}$ is the standard Poisson bracket $\{q,p\}_{_{\!\rm PB}}=1$, 
so $\{F,G\}_{_{\!\rm PB}}=
\frac{\partial F}{\partial q}
\frac{\partial G}{\partial p}-\frac{\partial G}{\partial q}
\frac{\partial F}{\partial p}$.

Of course, when the Hamiltonian function $H$ is time independent, $H(p,q)$ is itself  a solution. In what follows we assume that a solution for $K$ is known (possibly approximately
or even numerically). Then we make an ansatz for  the sought after diffeomorphism:
\begin{eqnarray}
\uppi&=& K(p,q,t)\, ,
\nonumber\\[1mm]
\upchi&=& -t + \phi(p,q,t)\, ,
\label{change}\\[1mm]
\uppsi &=& \lambda(p,q,t)\, .
\nonumber
\end{eqnarray}
Since ${\mathcal L}_\rho \uppi = 0 = {\mathcal L}_\rho{\mathcal \upchi}$ and ${\mathcal L}_\rho \uppsi=-1$, using Equation~\nn{yourboat} we must have 
$$
\frac{\partial K}{\partial t}=\{H,K\}_{_{\!\rm PB}}\, ,\quad
\frac{\partial \phi}{\partial t}=\{H,\phi\}_{_{\!\rm PB}}+1\, ,\quad
\frac{\partial \lambda}{\partial t}=\{H,\lambda,\}_{_{\!\rm PB}}+H-p\hh \frac{\partial H}{\partial p}\, .
$$
%
Comparing the right hand sides of Equations~\nn{alphaH} and~\nn{alphaD} and using the Ansatz~\nn{change}
gives a triplet of PDEs which we wish to use to determine $\phi$ and $\lambda$:
\begin{eqnarray}
K\hh\frac{\partial \phi}{\partial p}\,-\, \frac{\partial \lambda}{\partial p}&=&0\, ,
\nonumber\\[1mm]
K\hh\frac{\partial \phi}{\partial q}\hh -\frac{\partial \lambda}{\partial q}&=&p\, ,
\label{theeqs}
\\[1mm]
K\hh\frac{\partial \phi}{\partial t}-\frac{\partial \lambda}{\partial t}&=&K-H\, .\nonumber
\end{eqnarray}
Differentiating the second equation with respect to $p$ and the first
with respect to $q$ 
and then taking the difference yields
\begin{equation}\label{useme}
\{\phi,K\}_{_{\! PB}}=1\, .
\end{equation}
Assuming that the equation $\varepsilon = K(p,q,t)$ can be solved for $p=p(\varepsilon,q,t)$
we can solve the PDE given by the above display by first  writing 
$$
\phi(p,q,t)=:\upphi(K(p,q,t),q,t)\, .
$$
Using $\{K,K\}_{_{\! \rm PB}}=0$, 
Equation~\nn{useme} now says that
$$
\frac{\upphi(\varepsilon,q,t)}{\partial q}\frac{\partial K(p,q,t)}{\partial p}\Big|_{p=p(\varepsilon,q,t)}=1\, .
$$
Hence
$$
\upphi(\varepsilon,q,t)=\int^q
\frac{dx}{\, \frac{\partial K(p,x,t)}{\partial p}\, }
\!\!\!\!
\left.
\phantom{\frac{A}{B}\!\!}\right|_{p=p(\varepsilon,x,t)}
\, .
$$
Similarly, multiplying the 
second equation in Display~\nn{theeqs}
by $\partial K/\partial p$
and the first by $\partial K/\partial q$, the difference yields
$$
K\{\phi,K\}_{_{\! \rm PB}}
-\{\lambda,K\}_{_{\! \rm PB}}=p \frac{\partial K}{\partial p}\, .
$$
This can be solved for $\lambda(p,q,t)=:\uplambda(K(p,q,t),q,t)$ using the same method employed for~$\phi$:
$$
\uplambda(\varepsilon,q,t)=
\int^q dx\, 
\frac{\, \varepsilon- p\frac{\partial K(p,x,t)}{\partial p}\, }
{\frac{\partial K(p,x,t)}{\partial p}}
\!\!\!\!
\left.
\phantom{\frac{A}{B}\!\!}\right|_{p=p(\varepsilon,x,t)}
\, .
$$
In summary, the diffeomorphism to the Darboux coordinate system is given by
\begin{eqnarray}
\uppi&=& K(p,q,t)\, ,
\nonumber\\[2mm]
\upchi&=& -t + \int^q
\frac{dx}{\, \frac{\partial K(p,x,t)}{\partial p}\, }
\!\!\!\!
\left.
\phantom{\frac{A}{B}\!\!}\right|_{p=p(K(p,q,t),x,t)}\, ,
\label{solution}
\\[0mm]
\uppsi &=& 
\int^q dx\, 
\frac{\, K(p,q,t)- \Big(p\frac{\partial K(p,x,t)}{\partial p}\Big)\Big
|_{p=p(K(p,q,t),x,t)}\, }
{\frac{\partial K(p,x,t)}{\partial p}\Big
|_{p=p(K(p,q,t),x,t)}}
\, .\nonumber
\end{eqnarray}

\begin{example}\label{TISE}
For a time independent Hamiltonian
$$
H=\frac 12 \hh p^2 + \frac 12 \hh q^2 + v(q)
$$
we have
\begin{eqnarray}
\nonumber
\uppi &=&\, \, 
\frac12\hh  p^2 + \frac 12 \hh q^2 + v(q)\\[1mm]
\upchi &=&-t+
\int^q \frac{dx}{\sqrt{p^2+q^2-x^2+2(v(q)-v(x))}}\\
\label{covs}
&\approx&-t-\arctan\frac pq -\int^q dx\frac{v(q)-v(x)}
{(p^2+q^2-x^2)^{3/2}}\, ,
\nonumber
\\[2mm]
\nonumber
\uppsi&=&
\int^q dx\,  
\frac{x^2+2v(x)-\frac12 p^2 -\frac12 q^2-v(q)
}{\sqrt{p^2+q^2-x^2+2(v(q)-v(x))}}
\\
&\approx&
-\frac12\hh pq
-\int^q dx\, \frac{\frac12(p^2+q^2)(v(q)-3v(x))+x^2v(x)}
{(p^2+q^2-x^2)^{3/2}}
\, .
\end{eqnarray} 
The stated approximations
are accurate in the limit when the deformation of the harmonic oscillator $v(q)<\!\!<q^2$. 
In the harmonic oscillator limit when $v(q)=0$, it is easily checked that indeed
$$
\alpha=
\uppi d\upchi - d\uppsi = 
\frac12 (p^2+q^2) d\big(\!\!-t-\arctan \frac pq\big) 
-d\big(\!\!-\frac12 pq\big) = 
pdq - \frac12 (p^2 + q^2)dt\, .
$$
\end{example}

Now that we can explicitly locally map a wide class of classical dynamical systems to one another using the Darboux theorem, we proceed to study the quantum analog that was discussed in Section~\ref{QDsec}.
We shall focus on the quantum anharmonic oscillator.

\section{The Quantum Anharmonic Oscillator}

\noindent
Let us consider the  model with Hamiltonian
\begin{equation}\label{HAM}
H=\frac12\hh  p^2 + V(q)\, ,
\end{equation}
with a single-well potential $V$ that is smooth, concave up and obeys $V(0)=V'(0)=0
$. The classical phase-space curves for this model are concentric closed orbits. The phase-spacetime is ${\mathbb R}^3\ni (p,q,t)$ and
$$
\varphi=\big(dp + V'(q) dt\big)\wedge\big(dq- p dt\big)\, .
$$
 In the Darboux coordinates $(\upchi,\uppi,\uppsi)$, the physical trajectories are straight lines. Clearly for this model, these can be {\it globally} mapped to the 
trajectories in $(p,q,t)$ space. These are helix-like and foliate the phase-spacetime.

Standard quantization of the Hamiltonian~\nn{HAM} replaces the $c$-numbers $p$ and $q$ by quantum operators
$$
p\mapsto
\frac\hbar i \frac{\partial}{\partial S}\, ,\qquad
q\mapsto S\, ,
$$
acting on the Hilbert space ${\mathcal H}=L^2(\mathbb R)$ given by square integrable, complex-valued functions of $S$.
A common choice of quantum Hamiltonian is then
\begin{equation}\label{qHAM}
\widehat H := -\frac{\hbar^2}{2}
\frac{\partial^2}{\partial S^2}
+V(S)\, .
\end{equation}
In principle, the space of {\it all} quantizations of the classical Hamiltonian in Equation~\nn{HAM} ought be encoded in the space of flat quantum connections on the phase-spacetime. This issue is of independent interest, but we avoid studying it for now, and instead focus on the simple quantization in the above display.

Our aim is to compute the evolution operator $\exp\big(-\frac {i(t_f-t_i)}\hbar\widehat H \big)$
or its matrix elements
$$
K(S_f,S_i;t_f,t_i):=
\langle S_f|\exp\big(-\tfrac {i(t_f-t_i)}\hbar\widehat H\big)
|S_i\rangle\, . 
$$
This the usual  propagator problem in non-relativistic quantum mechanics, which can be handled perturbatively by various quantum mechanical techniques. Here we want to demonstrate a rather different approach based on the quantum Darboux theorem.

\medskip

First we need to rewrite the 
operator $\exp\big(-\frac i\hbar\widehat H (t_f-t_i)\big)$ as a path ordered line operator $P_\gamma \exp\big(-\int_\gamma \widehat A\hh \big)$ for a quantum connection $\nabla=d+\widehat A$  acting on  a Hilbert bundle over phase-spacetime. For that, we need a solution for $\nabla$ corresponding to the quantum Hamiltonian in Equation~\nn{qHAM}.
A solution is given by~\cite{Her}
\begin{equation}\label{AHOconn}
\nabla = d + \frac{dq}{i\hbar}
\Big(p+\frac{\hbar }{i}\frac{\partial}{\partial S}\Big)
-\frac{dp}{i\hbar}\, S
-\frac{dt}{i\hbar}\, \Big(\frac 12
\Big[p+\frac{\hbar}{i}
\frac{\partial}{\partial S}\Big]^2 + V(q+S)\Big)=:d+\widehat A\, .
\end{equation}
Observe that along the path
$$
\gamma = \{(0,q,t_i (1-\tau) + t_f\tau)\, :\, \tau\in [0,1]\}\in Z\, ,$$
the connection potential 
$$\widehat A= \frac i\hbar \widehat H_q (t_f-t_i)d\tau\, ,$$
where $\widehat H_q:=-\frac{\hbar^2}{2}
\frac{\partial^2}{\partial S^2}
+V(S+q)$.
It is tempting to take a path $\gamma$ along which both~$p$ and~$q$ vanish. However we avoid this choice because the Jacobian for a change of variables $(p,q)\mapsto (\uppi,\phi)$ vanishes along such a path. Because $\partial/\partial S=\partial/\partial (S+q)$, there is no difficulty  working along a path with constant $q\neq 0$, because this just shifts the variable~$S$ in wavefunctions.

The soldering forms for the above connection are given by
$$
e^a=\big(dp+V'(q)dt,dq-p dt\big)=:(f,e)\, ,
$$
so that $\varphi= \frac12 j_{ab} e^a\wedge e^b=e\wedge f$ and $j_{12}=1=-j_{21}=j^{12}=-j^{21} $.
These forms vanish along $\gamma$. 
Also note that $\hat s_a=\big(-S,\frac\hbar i \frac{\partial}{\partial S}\big)$ and $[\hat s_1,\hat s_2]=-i\hbar$.
It now follows that
\begin{equation}\label{woe}
P_\gamma \exp\Big(-\int_\gamma \widehat A\hh \Big)
=
\exp\Big(
-\frac {i(t_f-t_i)
} \hbar \int_0^1 d\tau\, 
\widehat H_q \Big)
=
\exp\Big(-\frac i\hbar\widehat H_q (t_f-t_i)\Big)\ .
\end{equation}
Our next goal is to find a gauge transformation  mapping (at least formally) the connection~$\nabla$ to a far simpler one. Before doing that, it useful to discuss quantum symmetries of this system.

\subsection{Quantum Noether Theorem}

Since quantum dynamics is given by parallel transport with respect to a quantum connection $\nabla=d+\widehat A$, quantum symmetries ought be given by operators $\widehat {\mathcal O}$ on the Hilbert bundle that obey
$$
\nabla \circ \widehat {\mathcal O}=\widehat {\mathcal O}{}\hh ^\prime\circ \nabla \, ,
$$
for any operator $\widehat{\mathcal O}{}\hh ^\prime$, since if $\Psi\in \Gamma({\mathcal H}Z)$
solves $\nabla \Psi =0$, then so too does $ \widehat {\mathcal O}\Psi$.
\smallskip

Such operators $\widehat {\mathcal O}$ are easy to construct:
Let $u\in \Gamma(TZ)$ be a vector field.
Then, because $\nabla$ is nilpotent,  the operator
$\{\iota_u,\nabla\}$ commutes with $\nabla$. Acting on sections of the Hilbert bundle, this gives the operator
$$
{\mathcal L}_u + \widehat A_u\, .
$$
Note, that the Lie derivative is defined acting on sections of $\wedge^\bullet Z\otimes {\mathcal H}Z$ is defined by the anticommutator~$\{\iota_u,d\}$.
Symmetries of the above  type that hold for arbitrary vector fields $u$ are tautological, in the sense that 
the first operator acts along the base $Z$, while the second acts on the  Hilbert space fibers in a way that exactly compensates the former transformation. In BRST terms, these symmetries are  BRST exact. However, specializing to vector fields $u$ that solve
 \begin{equation}
 \label{qsymm}
 [{\mathcal L}_u ,\nabla] = \frac{d\beta}{i\hbar}
 \end{equation}
 for some $\beta \in C^\infty Z$,
 we can define a {\it conserved quantum charge}
 \begin{equation}\label{chargemaker}
 \widehat Q_u:=i\hbar \{\iota_u,\nabla\}-i\hbar {\mathcal L}_u-\beta\, .
 \end{equation}
 It is easy to verify that
 $$
 [ \nabla,\widehat Q_u ]=0\, .
 $$
 Note that the leading term of $\widehat Q_u$ in the grading is $\alpha(u)-\beta$
which gives a classical conserved charge $Q$ that solves Equation~\nn{clQ} because
$$
{\mathcal L}_\rho Q = \iota_\rho d (\iota_u \alpha) -\iota_\rho d\beta=\iota_\rho {\mathcal L}_u\alpha-\iota_\rho\iota_u \varphi-\iota_\rho d\beta=0\, .
$$
Here we used that Equation~\nn{qsymm} implies
${\mathcal L}_u \alpha = d\beta$ and $\varphi(\rho,\cdot)=0$.
 Let us call vector fields $u$ obeying Equation~\nn{qsymm} {\it quantum symmetries} of $\nabla$. 
 
 \smallskip
The simple model in the next example 
will be important.

\begin{example}
Let $Z={\mathbb R}^3\ni (\uppi,\upchi,\uppsi)$ and 
$$
\varphi=d\uppi\wedge d\upchi\, .$$ 
This equals $d\alpha$ where $\alpha=\uppi d\upchi-d\uppsi$ is contact and the coordinates $(\uppi,\upchi,\uppsi)$ are Darboux.
Then a quantum connection is
\begin{equation}\label{AHOD}
\nabla_{\rm D} = d + \frac{d\upchi}{i\hbar}
\Big(\uppi+\frac{\hbar }{i}\frac{\partial}{\partial S}\Big)
-\frac{d\uppi}{i\hbar}\, S
-\frac{d\uppsi}{i\hbar} =:d+\widehat A_{\rm D}\, .
\end{equation}
The vector fields 
$$
\frac{\partial }{\partial \uppi}\, ,\quad
\frac{\partial }{\partial \upchi}\, ,\quad
\frac{\partial }{\partial \uppsi}\, ,
$$
are quantum symmetries with $\beta$ equaling $\upchi$, $0$, $0$, respectively. Their conserved quantum charges are
\begin{equation}\label{Darcha}
\widehat Q_{\frac{\partial }{\partial \uppi}}=-\upchi-S\, ,\quad
\widehat Q_{\frac{\partial }{\partial \upchi}}=\uppi+\frac{\hbar }{i}\frac{\partial}{\partial S}\, ,\quad
\widehat Q_{\frac{\partial }{\partial \uppsi}}=-1\, .\end{equation}
Acting on the Hilbert bundle, these charges obey the Heisenberg Lie algebra which is also the algebra of the three contact Hamiltonian vector fields $\frac{\partial }{\partial \uppi}+\upchi\frac{\partial }{\partial \uppsi}$,
$\frac{\partial }{\partial \upchi}$ and
$\frac{\partial }{\partial \uppsi}$.
\end{example}

Returning to the anharmonic oscillator, we note  that it has two  non-trivial,  independent, globally defined, conserved charges. One of these is the Hamiltonian $H=\frac12 p^2 + V(q)$, or simply~$\uppi$ in Darboux coordinates. The other is $t-\phi(p,q)$, or equivalently~$-\upchi$. This says that the angle variable (see Equation~\nn{solution})
\begin{equation}\label{angle}
\phi(p,q)=\int^q\frac{dx}{\sqrt{p^2+2V(q)-2V(x)}}\, ,
\end{equation}
obeys $\phi(p,q)=t+{\rm constant}$. Or in other words, the initial value of the angle variable is preserved along classical paths.

It is interesting to study the quantization of the conserved quantities $H$ and $t-\phi$. The former is simple because
$$
[{\mathcal  L}_{\frac{\partial}{\partial t}},\nabla]=0
$$ 
for the quantum connection form given in Equation~\nn{AHOconn}. Hence we can construct the quantum charge corresponding to the strict contactomorphism generated by $\frac{\partial}{\partial t}$. This gives
$$
-\widehat Q_{\frac{\partial}{\partial t}}
=\frac 12
\Big[p+\frac{\hbar}{i}
\frac{\partial}{\partial S}\Big]^2 + V(q+S)\, .
$$
It is easy to verify that the above operator commutes with $\nabla$. Moreover, at $p=0=q$, this recovers the standard quantum Hamiltonian $\widehat H$.

The quantum charge corresponding to $t-\phi(p,q)$ is more involved. The quantization given by the quantum connection form $\nabla$ in Equation~\nn{AHOconn} does not obey  Condition~\nn{qsymm} for the vector field $u=
 \frac{\partial }{\partial \uppi}+\upchi\frac{\partial }{\partial \uppsi}$.
 This does not mean that there is no corresponding charge, but rather the simple Formula~\nn{chargemaker} cannot be used. Instead, given  a quantum gauge transformation $\hat U$ relating $\nabla$ to $\nabla_{\rm D}$
 as in Equation~\nn{QU}, 
 then the  quantum charge~$
 \hat U^{-1} \widehat Q_{\frac{\partial}{\partial \uppi} }\hat U
 $ commutes with $\nabla$. Unfortunately we do not yet know the gauge transformation $\hat U$. The computation of this operator is the subject of the next section.

%
%

\subsection{Quantum Gauge Transformation}

We want to compute the quantum gauge transformation $\hat U$ given by a unitary endomorphism of the section space of the Hilbert bundle $\Gamma({\mathcal H}Z)$
such that
$$
\nabla=\hat U^{-1} \nabla_{\rm D} \hat U\, ,
$$
where $\nabla$ is the connection form corresponding to the anharmonic oscillator in Equation~\nn{AHOconn}
while $\nabla_{\rm D}$ is the Darboux connection in Equation~\nn{AHOD}.
For this we will work iteratively order by order in the grading. The first step is to find a gauge transformation relating  the calibration maps of the two connection forms. 

\subsubsection{Metaplectic transformation}

To begin with, we need the relation between the soldering forms $(d\uppi,d\upchi)$ of the  Darboux connection $\nabla_{\rm D}$  to those---given by $(dp+H_q dt, dq-H_p)$---of~$\nabla$. (For brevity, from now on we often denote partial derivatives by subscripts). This is given by
$$
\begin{pmatrix}d\uppi\\ d\upchi
\end{pmatrix}
=\frac{\partial(\uppi, \phi)}{\partial(p,q)}
\begin{pmatrix}dp\\ dq
\end{pmatrix}
-dt \begin{pmatrix}0\\1\end{pmatrix}
=\begin{pmatrix}
\uppi_p & \uppi_q\\
\phi_p & \phi_q
\end{pmatrix}
\begin{pmatrix}dp+H_q dt\\ dq-H_pdt
\end{pmatrix}\, .
$$
The above display was computed using Equation~\nn{change}
specialized to the case $K=H(p,q)$ for which the functions $\phi$ and $\lambda$ are $t$-independent. In the  above
\begin{equation}\label{V}
U_0=\frac{\partial(\uppi, \phi)}{\partial(p,q)}=\begin{pmatrix}
\uppi_p & \uppi_q\\
\phi_p & \phi_q
\end{pmatrix}
\end{equation}
is the Jacobian of the change of variables $(p,q)\to (\uppi,\phi)$. Notice also, that the last equality was achieved using that the Poisson bracket of $\{\uppi,H\}_{_{\! \rm PB}}=0$ and $\{\phi,H\}_{_{\! \rm PB}}=1$ (see Equation~\nn{useme}).
This implies that
$$\det U_0=1\, .$$
Hence the matrix $U_0$ is $Sp(2)$-valued with respect to the antisymmetric bilinear form 
$$
J=\begin{pmatrix}
0&1\\-1&0
\end{pmatrix}=:(j_{ab})\, .
$$
We want to intertwine the $Sp(2)$ group element $U_0$ expressed in the fundamental representation
in Equation~\nn{V} to an operator  $\hat U_0$ acting on sections of the Hilbert bundle. 
This operator must obey 
$$
\hat U_0^{-1}\Big(\frac{d\upchi}{i\hbar}
\frac{\hbar }{i}\frac{\partial}{\partial S}
-\frac{d\uppi}{i\hbar}\, S
\Big)\hat U_0 =
\frac{dq-H_p dt}{i\hbar}
\frac{\hbar }{i}\frac{\partial}{\partial S}
-\frac{dp + H_q dt}{i\hbar}\, S
\, .
$$
{\it I.e.}, the operator $\hat U_0$ transforms the Darboux solderings to those of $\nabla$.
The  map from the matrix $U_0$ to the operator $\hat U_0$ is  the intertwiner from the fundamental representation of $Sp(2)$ to its metaplectic representation on the projective Hilbert space ${\mathbb P}({\mathcal H})$. Note that strictly only the double cover $Mp(2)$ of $Sp(2)$  has a metaplectic representation on the Hilbert space, but upon projectivizing, this gives a unitary $Sp(2)$ representation. Formally, the {\it projective} metaplectic action of any $Sp(2)$ matrix $V$ is determined by the formul\ae\
$$
\widehat{\scalebox{.9}{$\begin{pmatrix}1&0\\ t&1\end{pmatrix}$}}=
\exp\Big(
\frac{it\hbar }{2}  \frac{\partial ^2}{\partial S^2}
\Big)\, ,\quad
\widehat{\scalebox{.9}{$\!\begin{pmatrix}e^{-\ell}&0\\0&e^{\ell}\end{pmatrix}\!$}}=
\exp\Big(-\ell S\frac{\partial}{\partial S}\Big)\, , \quad
\widehat{\scalebox{.9}{$\begin{pmatrix}1&u\\ 0&1\end{pmatrix}$}}=
\exp\Big(
\frac{iu}{2\hbar} S^2
\Big)\, .
$$
The action of these operators on wavefunctions $\psi(S)$ can be computed using suitable Fourier transforms. In particular (up to irrelevant normalizations)
\begin{equation}\label{letsgetmeta}
\,\,
\widehat{\scalebox{.9}{$\begin{pmatrix}1&0\\t&1\end{pmatrix}$}}\psi(S)\!=\!
\int dS' e^{\frac {i(S-S')^2}{2\hbar t}}\!\psi(S')\, ,\quad\!\!\!\!\!
\widehat{\scalebox{.9}{$\!\begin{pmatrix}e^{-\ell}\!&0\\0&\!e^{\ell}\end{pmatrix}\!$}}\, \psi(S)=\psi(e^{-\ell} S)\, , \quad\!\!\!\!\!
\widehat{\scalebox{.9}{$\begin{pmatrix}1&u\\0&1\end{pmatrix}$}} \psi(S) = e^{
\frac{iu}{2\hbar} S^2}\!\psi(S). 
\end{equation}
Also, again up to an irrelevant normalization,
\begin{equation}\label{Fourier}
\widehat{\scalebox{.9}{$\!\begin{pmatrix}0\!&\!1\\-1\!&\!0\end{pmatrix}\!$}}\, 
\psi(S)=\int dS' e^{-\frac i\hbar S S'} f(S')\, .\end{equation}

For future use, note that at the level of  the Lie algebra $ \mathfrak{sp}(2)$---recycling the hat notation for this---one has  
\begin{equation}\label{Idonotlie}
\widehat {\begin{pmatrix}-a&b \\c&a\end{pmatrix}}=
\frac{i\hbar c}{2}\, \frac{\partial^2}{\partial S^2}
-a\Big(S\frac{\partial}{\partial S}
+\frac12\Big)
+
\frac{ib}{2\hbar}\, S^2
=\frac1{2!i\hbar} M^{ab} \hat s_a \hat s_b\, .
\end{equation}
In the above $\big(M^a{}_b\big):=\begin{pmatrix}-a&b \\c&a\end{pmatrix}$ and $M^a{}_b=:j_{bc}M^{ac}$.

Also note that if $AD-BC=1$, then
\begin{equation}\label{letsgetmeta}
\widehat{\begin{pmatrix}
A & B \\C & D
\end{pmatrix}}
\circ\Big(-\alpha S + \beta \frac\hbar i \frac{\partial}{\partial S}
\Big)
\circ
\widehat{\begin{pmatrix}
\!D\! & \!\!-B\! \\\!\!-C\! & \!A\!
\end{pmatrix}}
=
-[A\alpha +B \beta] S + [C\alpha + D \beta]\, \frac\hbar i \frac{\partial}{\partial S}\, .
\end{equation}
The above formula is the
  intertwiner between the fundamental representation of $Sp(2)$ and its projective metaplectic representation.

By now we have achieved that
$$
\hat U_0^{-1} \nabla_{\rm D} \hat U_0=
\frac{pdq-Hdt}{i\hbar}+
\frac{dq-H_p dt}{i\hbar}
\frac{\hbar }{i}\frac{\partial}{\partial S}
-\frac{dp + H_q dt}{i\hbar}\, S
+
\widehat{
\scalebox{.7}{$
\begin{pmatrix}
\phi_q \!& \!-\uppi_q\\
-\phi_p \!& \!\uppi_p
\end{pmatrix}$}}\circ d\circ
\widehat{
\scalebox{.7}{$\begin{pmatrix}
\uppi_p & \uppi_q\\
\phi_p & \phi_q
\end{pmatrix}$}}\, .
$$
The difference between $\nabla_0$ and $\nabla$ in Equation~\nn{a}
is given by (see also Equation~\nn{AHOconn})
$$
\hat a_0 = 
\widehat{
\scalebox{.7}{$
\begin{pmatrix}
\phi_q \!& \!-\uppi_q\\
-\phi_p \!& \!\uppi_p
\end{pmatrix}$}}
 \:\Big(d
  \widehat{
\scalebox{.7}{$\begin{pmatrix}
\uppi_p & \uppi_q\\
\phi_p & \phi_q
\end{pmatrix}$}}
\Big)
+\frac{dt}{i\hbar}\, \Big(-\frac {\hbar^2}2
\frac{\partial^2}{\partial S^2} + v_2(q,S)\Big)\, .
$$
In the above $v_2:=V(q+S)-V(q)-V'(q)S$, and the above-displayed operator only has terms of grade zero and higher.

To complete the computation of $\hat a_0$, we must calculate $\hat U_0^{-1} d\hat U_0$.
Because this one-form is Lie algebra-valued, we instead compute~$U_0^{-1}dU_0$. Note that the matrix $U_0$ in Equation~\nn{V} only depends on the variables~$p$ and~$q$, or equivalently only on the pair $(\uppi,\phi)$. Moreover, $\upchi=-t+\phi$, so observe that---acting on functions that depend only on $(p,q)$---the exterior derivative can be written
$$
d = d\uppi \frac{\partial}{\partial  \uppi} + (d\upchi+dt)\frac{\partial}{\partial  \phi}\, .
$$ 
Moreover
$$
\frac{\partial}{\partial \pi}=\phi_q\partial_p-\phi_p\partial_q=\{\phi,\cdot\}_{_{\! \rm PB}}
\,, \qquad
\frac{\partial}{\partial \phi}
=
-V'\partial_p+p\partial_q
=-\{\uppi,\cdot\}_{_{\! \rm PB}}
\, .
$$
Then, using $\{\phi,\uppi\}_{_{\! \rm PB}}=1$, after some computation, it follows that
\begin{equation}\label{jhess}
U_0^{-1} \frac{\partial}{\partial \uppi}
U_0=\begin{pmatrix}
-\phi_{pq}&-\phi_{qq}\\
\phi_{pp}& \phi_{pq}
\end{pmatrix}= -J\, \operatorname{Hess}( \phi)
\, ,\:\:
U_0^{-1} \frac{\partial}{\partial \phi}
U_0= 
\begin{pmatrix}
\uppi_{pq}&\uppi_{qq}\\
-\uppi_{pp}& -\uppi_{pq}
\end{pmatrix}
=J\, \operatorname{Hess}( \uppi)
\, ,
\end{equation}
where the Hessian matrix $\operatorname{Hess}(f):=\begin{pmatrix} f_{pp}&f_{pq}\\f_{pq} & f_{qq}\end{pmatrix}$.   Because the Hessian is symmetric, it follows that $J\operatorname{Hess}(f) \in {\mathfrak {sp}}(2)$. 

\medskip 

Using Equation~\nn{Idonotlie}, orchestrating the above computations gives
$$
\hat a_0 = 
\widehat{\begin{pmatrix}
0\!&\!V''\\
-1\!&\! 0
\end{pmatrix}} d\upchi +
\widehat{
\begin{pmatrix}
\!-\phi_{pq}\!\!&\!\!-\phi_{qq}\!\\
\!\phi_{pp}\!\!& \!\!\phi_{pq}\!
\end{pmatrix}}
d\uppi
+
\frac{dt}{i\hbar}\, v_3(q,S)
\, ,$$
where $v_3:=V(q+S)-V(q)-V'(q)S-\frac1{2!}V''(q)S^2$
and the angle variable is given explicitly in Equation~\nn{angle}.
Happily---and necessarily on general grounds---the first two terms in the above display have grade zero, and lie in the codistribution. Assuming real analyticity of $V(S)$, the last term has grades one and higher.
Next we need to compute the higher order gauge transformation $\hat U_1$ subject to Equation~\nn{Schrod}. 
\bigskip

\subsubsection{Higher order gauge transformations}

To compute $\hat U_1$ we work in a formal power series in the grading. Examining~\nn{Schrod}, we see that it is simpler to compute $\hat U_1^{-1}$, which we expand as
$$
\hat U_1^{-1}=1
+\frac{W^{abc}\hat s_a \hat s_b \hat s_c}{3!i\hbar}+\cdots\, .
$$
Here $W^{abc}$ is some totally symmetric tensor to be determined.
Then the lowest order contribution to Equation~\nn{Schrod} implies that
\begin{equation}\label{Ahi}
\left[\frac{e^a\hat s_a}{i\hbar}
,\frac{W^{bcd}\hat s_b \hat s_c \hat s_d}{3!i\hbar}\right]=
\frac{e^a j_{ab} W^{bcd} \hat s_c \hat s_d}{2!i\hbar}
=
\widehat{\begin{pmatrix}
0\!&\!V''\\
-1\!&\! 0
\end{pmatrix}} d\uppi +
\widehat{
\begin{pmatrix}
\!-\phi_{pq}\!\!&\!\!-\phi_{qq}\!\\
\!\phi_{pp}\!\!& \!\!\phi_{pq}\!
\end{pmatrix}}
d\upchi\, .
\end{equation}
Using Equations~(\ref{Idonotlie},\ref{jhess}) and calling $\Pi^a = (\uppi, \phi)$ and $\partial_a=(\partial_p,\partial_q)$ we have that
$$
W_{abc}=j_{fe}\partial_a \Pi^e \, \partial_b\partial_c \Pi^f\, .
$$
Moreover, using that $\partial_{[a} \Pi^e \, \partial_{b]} \Pi^f$ is proportional to the Poisson bracket of $\Pi^e$ and $\Pi^f$, and that $\{\phi,\uppi\}_{_{\! \rm PB}}
=1$, it follows that $W_{abc}$ is totally symmetric. 

Now let us examine higher order terms in $\hat U_1^{-1}$. Let us call the grade 1 object 
\begin{equation}\label{What1}
\hat W_{(1)}:=\frac{W^{abc}\hat s_a \hat s_b \hat s_c}{3!i\hbar}
\end{equation} and search for the grade 2 correction
$$
\hat U^{-1}_1=1+\hat W_{(1)}+
\hat W_{(2)}+\cdots\, .
$$
Also, let us decompose
$$\hat a_0=a_{(0)}+a_{(1)}+\cdots\, ,$$
where 
$$\hat a_{(0)} = 
\widehat{\begin{pmatrix}
0\!&\!V''\\
-1\!&\! 0
\end{pmatrix}} d\upchi +
\widehat{
\begin{pmatrix}
\!-\phi_{pq}\!\!&\!\!-\phi_{qq}\!\\
\!\phi_{pp}\!\!& \!\!\phi_{pq}\!
\end{pmatrix}}
d\uppi
\:
\mbox{ and }\:\hat a_{(1)}
=
\frac{dt}{3!i\hbar}\, V'''(q)S^3\, .
$$
Along, similar lines, the grade zero part of $\nabla$ in Equation~\nn{AHOconn} is
$$
\nabla_{(0)}:=
 d 
-\frac{dt}{i\hbar}\, \Big(-\frac {\hbar^2}2
\frac{\partial^2}{\partial S^2} +
\frac1{2} V''(q)S^2\Big)\, .
$$
Then, at grade 1, Equation~\nn{Schrod} demands that
\begin{equation}\label{Fugu}
\left[\frac{e^a\hat s_a}{i\hbar}
,\hat W_{(2)}\right]=-
\big[\nabla_{(0)},\hat W_{(1)}\big]
+\hat a_{(1)}+
\hat W_{(1)} \hat a_{(0)}\, .
\end{equation}

To solve the above equation it is important to remember that $\hat U_1^{-1}$ must be  a unitary operator. Thus we must require
\begin{equation}\label{gettinghe}
(\hat U^{-1})^{-1} = 1-\hat W_{(1)}
-\hat W_{(2)}+ \hat W_{(1)}^{\, 2}\cdots
=
1+\hat W_{(1)}^\dagger+
\hat W_{(2)}^\dagger+\cdots=
(\hat U^{-1})^{\dagger}\, ,
\end{equation}
so the hermitean part of $\hat W_{(2)}$ is given by
$$
\operatorname{He}(\hat W_{(2)})=\frac12\, \hat W_{(1)}^{\, 2}\, .
$$
The anti-Hermitean part $\operatorname{aHe}(W_{(2)})$
is still undetermined. However  Equation~\nn{Ahi} says that
$
\left[\frac{e^a\hat s_a}{i\hbar}
,\hat W_{(1)}\right]=\hat a_{(0)}
$
so Equation~\nn{Fugu} then gives
\begin{equation}\label{rhs}
\left[\frac{e^a\hat s_a}{i\hbar}
,\operatorname{aHe}(\hat W_{(2)})\right]=-
\big[\nabla_{(0)}
+\frac12\,  \hat a_{(0)}
,\hat W_{(1)}\big]
+\hat a_{(1)}\, .
\end{equation}
To proceed we need to compute the right hand side of the above expression. This is tractable computation, but clearly as we move to even higher orders the complexity of such computations will grow dramatically. Hence, we digress to develop a diagrammatic representation of the operator differential forms appearing in the above discussion.

\subsubsection{Heaven and Earth Diagrams}\label{H+E}

Recall that $\Pi^a:=(\uppi,\phi)$ and $\partial_a:=(\partial_p,\partial_q)$. Let use depict the tensor obtained from partial derivatives on $\Pi^a$ by
$$
\partial_{a_1}\partial_{a_2}\cdots \partial_{a_n} \Pi^b :=
\raisebox{-.5cm}{\includegraphics[width=2cm]{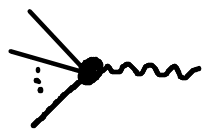}}
\, .
$$
\noindent
Here the solid lines correspond to the indices $a_1,\ldots,a_n$ and these may be permuted at no cost because partial derivatives commute.
The tensors $j_{ab}$ and its inverse $j^{ab}$ 
are denoted by directed  line segments
$$
j_{ab} :=
\raisebox{-3.5mm}{\includegraphics[width=2cm]{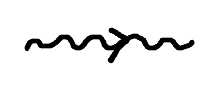}}
\mbox{ and }
j^{ab} :=
\raisebox{-3.1mm}{\includegraphics[width=2cm]{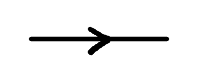}}\, .
$$
Concatenated lines indicated index contractions so, for example,
$$
\partial_{a_1}\partial_{a_2}\cdots \partial_{a_n} \Pi^b j_{bc} = \raisebox{-4.5mm}{\includegraphics[width=3.3cm]{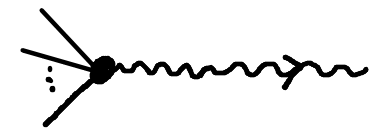}}\, ,
$$
and 
\begin{equation}\label{identity}
\raisebox{-1.5mm}{\includegraphics[width=2.2cm]{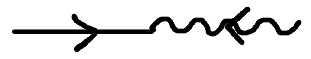}}=j_{ab}j^{cb}=\delta_a^c\, .
\end{equation}
The identity above allows the depicted concatenation to be removed from a diagram.
Also, note that reversing the direction of any arrow multiplies the tensor depicted by a minus sign.

More complicated diagrams obtained by such concatenations have the drawback that symmetry of external legs may be broken, and thus it may no longer be possible to uniquely associate a tensor to such a picture.
Moreover, we are typically interested in Hilbert space operators taking values in differential forms built from tensors made from derivatives and products of $\Pi^a$'s.
Hence we adopt a ``heaven and earth notation'' in which a line attached to the earth (a horizontal green line) denotes contraction with the operator $\hat s^a$. Similarly, 
a  line attached to heaven  denotes contraction with a differential form, the choice of which  will be labeled when this is not clear. For example
$$
\hat s^a \hat s^b \hat s^c \, \partial_a \partial_b \Pi^d j_{de} \partial_c \Pi^e= \raisebox{-4.5mm}{\includegraphics[width=4cm]{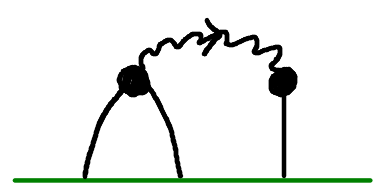}}\, .
$$
and
$$
\hat s^a \hat s^b \partial_a \partial_b \Pi^c j_{cd} \, d\Pi^d=\:\:
\stackrel{d\Pi}{
\raisebox{-11mm}{\includegraphics[width=3cm]{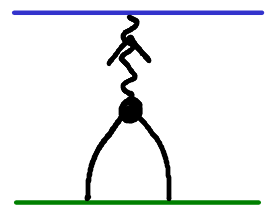}}}\, .
$$

Now the Poisson bracket $\{\phi,\pi\}=1$ implies that $\partial_a \Pi^c j_{cd} \partial_b \Pi^d = j_{ba}$ so we have the diagrammatic identity
\begin{equation}\label{poissa}
\raisebox{-2mm}{{\includegraphics[width=3cm]{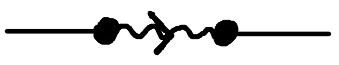}}}
=
\raisebox{-2.5mm}{{\includegraphics[width=1.7cm]{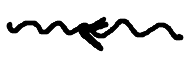}}}\,\,  .
\end{equation}
Similarly $\partial_a \Pi^c j^{ab} \partial_b \Pi^d = j^{cd}$
implies
\begin{equation}\label{poissb}
\raisebox{-2mm}{{\includegraphics[width=3cm]{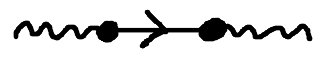}}}
=
\raisebox{-3mm}{{\includegraphics[width=2cm]{j2}}}.
\end{equation}
These relations imply two identities for heaven and earth diagrams
$$
\raisebox{-3mm}{{\includegraphics[width=3cm]{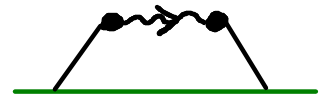}}}
=
\raisebox{-3.5mm}{{\includegraphics[width=2.4cm]{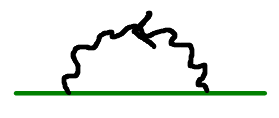}}}\, 
=i\hbar\, .
$$
\smallskip
$$
\stackrel{d\Pi\qquad \quad \:\; \:d\Pi}{
\raisebox{-6mm}{{\includegraphics[width=3cm]{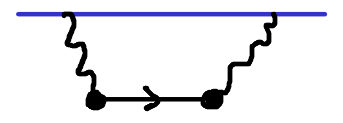}}}}
=
\stackrel{d\Pi\qquad d\Pi}{
\raisebox{-5.5mm}{{\includegraphics[width=2.4cm]{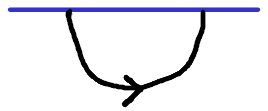}}}}
\, =-2\varphi\, .
$$
The last equalities used that $[\hat s^a,\hat s^b]=i\hbar j^{ba}$ and $j_{ab}d\Pi^a \wedge d\Pi^b = 2\varphi$.

\smallskip
Since a solid line attached to a solid dot denotes a partial derivative of $\Pi$, there are further identities following from the product rule of, for example,  the schematic type $(\partial \Pi) (\partial^2 \Pi) (\partial^3 \Pi) = \partial \big((\partial \Pi)^2 (\partial^3 \Pi)\big) -(\partial^2 \Pi) (\partial \Pi) (\partial^3 \Pi)-(\partial \Pi)^2 (\partial^4 \Pi) $.
Hence, in a heaven and earth diagram, we can remove a leg from a solid dot and produce (minus) a sum of diagrams with that leg attached to all other solid dots plus a term where this derivative acts on the whole diagram. The latter is denoted by a solid line that ends midair between heaven and earth and which denotes a partial derivative acting to the right. An example is the following identity
$$
\raisebox{-4.5mm}{\includegraphics[width=3.3cm]{Earth}}
=
\raisebox{-4.7mm}{\includegraphics[width=3.3cm]{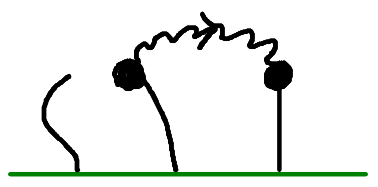}}
-
\raisebox{-4.5mm}{\includegraphics[width=3.3cm]{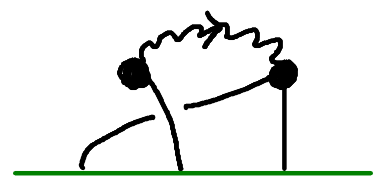}}
=
\raisebox{-4.5mm}{\includegraphics[width=3.3cm]{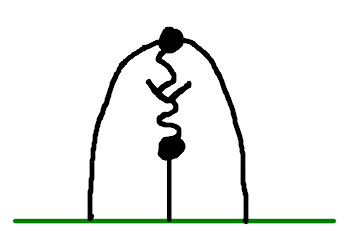}}\, .
$$
The last equality used that $\hat s^a \partial_a i\hbar =0$.
The above is a diagrammatic demonstration that the tensor $W_{abc}=j_{fe}\partial_a \Pi^e \, \partial_b\partial_c \Pi^f$ is totally symmetric. To emphasize this we diagrammatically denote 
$$
\raisebox{-7.6mm}{\includegraphics[width=3.5cm]{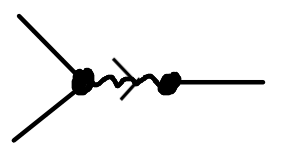}}
:=
\raisebox{-7.3mm}{\includegraphics[width=2.5cm]{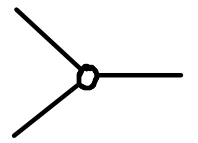}}\, .
$$

We need one further ingredient to attack Equation~\nn{rhs}; the exterior derivative. For that we note that acting on functions that are independent of $t$, we have
$$
d = d \uppi \hh \partial_\uppi + d\phi \partial_\phi = 
d \uppi\hh \{\phi,\cdot\}_{_{\! \rm PB}}
- d \phi \{\uppi,\cdot\}_{_{\! \rm PB}}
=
d\Pi^a j_{ba} \partial_c \Pi^b j^{cd} \partial_d\, .
$$
Thus we have the diagrammatic notation for the exterior 
derivative
$$
d:=\:\stackrel{d\Pi}{\raisebox{-8.4mm}{\includegraphics[width=2.5cm]{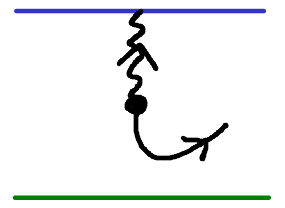}}}\, .
$$
Before proceeding, we need one further diagrammatic identity expressing
that $[\hat s^a,\hat s^b]=i\hbar j^{ba}$, namely
\begin{equation}\label{racoon-drone}
\raisebox{-5mm}{\includegraphics[width=2.5cm]{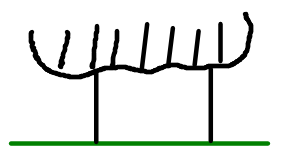}}
=
\raisebox{-5mm}{\includegraphics[width=3cm]{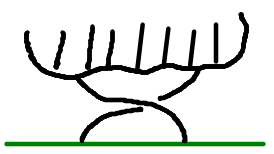}}
+i\hbar 
\raisebox{-5mm}{\includegraphics[width=2.5cm]{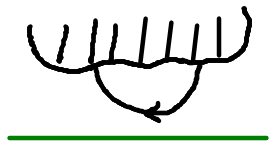}}\, ,
\end{equation}
where the shaded blobs denote the (unspecified) remainder of the diagram.

Now we are ready to diagrammatically compute the right hand side of Equation~\nn{rhs}. We begin with $[d,\hat W_{(1)}]$:
$$
\stackrel{\hspace{-2cm}d\Pi}{
\raisebox{-7mm}{\includegraphics[width=3.3cm]{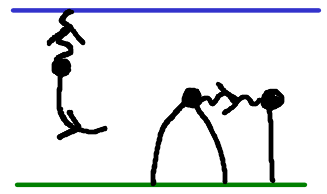}}
}
=
\stackrel{\hspace{-1cm}d\Pi}{
\raisebox{-7mm}{\includegraphics[width=2.2cm]{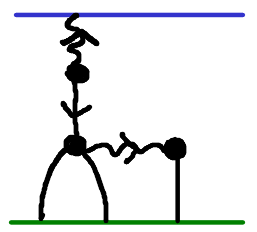}}}
+
\stackrel{\hspace{1cm}d\Pi}{
\raisebox{-6.6mm}{\includegraphics[width=2cm]{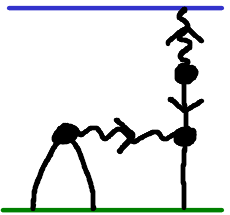}}}\, .
$$
Now let $\Omega^a$ be any pair of differential forms.  Then we
have the following identities
\begin{eqnarray*}
 \stackrel{\hspace{-1cm}\Omega}{
\raisebox{-9mm}{\includegraphics[width=2.2cm]{dW1}}}
& = & 
  \stackrel{\hspace{-.7cm}\Omega}{
\raisebox{-9mm}{\includegraphics[width=2.6cm]{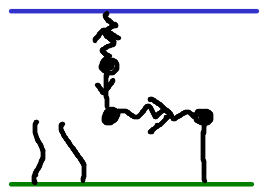}}}
-
  \stackrel{\hspace{-.7cm}\Omega}{
\raisebox{-9mm}{\includegraphics[width=2.5cm]{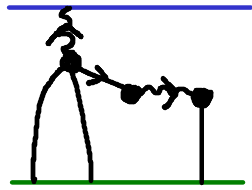}}}
-
  \stackrel{\hspace{-.7cm}\Omega}{
\raisebox{-9mm}{\includegraphics[width=2.6cm]{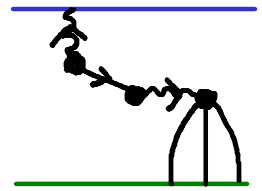}}}
-( \mbox{6  terms})
\\[3mm]
&=&
 \stackrel{\hspace{0cm}\Omega}{\raisebox{-9mm}{\includegraphics[width=2.6cm]{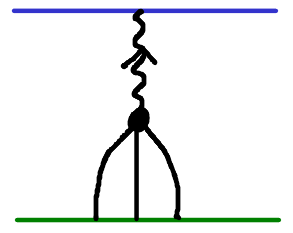}}}
-( \mbox{6  terms})
\, ,\\[1mm]
&=& \stackrel{\Omega}{\raisebox{-9mm}{\includegraphics[width=2.6cm]{Pilat}}}
+2
\stackrel{\hspace{1.5cm}\Omega}{
\raisebox{-9mm}{\includegraphics[width=2cm]{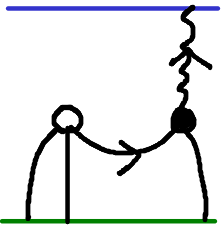}}}
+i\hbar
\stackrel{\hspace{-.2cm}\Omega}{
\raisebox{-9.1mm}{\includegraphics[width=1.8cm]{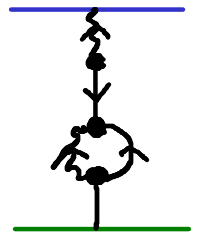}}}
\end{eqnarray*}
Hence 
$$
\stackrel{\hspace{-2cm}d\Pi}{
\raisebox{-8.5mm}{\includegraphics[width=3.4cm]{dW}}
}
=
\stackrel{\hspace{-0cm}d\Pi}{
\raisebox{-9mm}{\includegraphics[width=2.6cm]{Pilat}}}
+3
\stackrel{\hspace{1.5cm}d\Pi}{
\raisebox{-8.8mm}{\includegraphics[width=2cm]{pope2}}}
+3i\hbar
\stackrel{\hspace{-.2cm}d\Pi}{
\raisebox{-9mm}{\includegraphics[width=1.8cm]{PearlHarbor}}}
$$
Here we used that $\raisebox{-1mm}{\includegraphics[width=2.5cm]{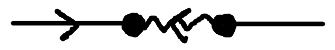}}=
\raisebox{-.5mm}{\includegraphics[width=1.8cm]{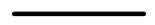}}$ and
$\raisebox{-1mm}{\includegraphics[width=2.1cm]{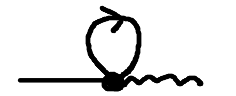}}=
0$.
For the remaining commutator terms in Equation~\nn{rhs} we need the following computation
\begin{eqnarray*}
\left[
\stackrel{\hspace{-2cm}\Omega}{
\raisebox{-9mm}{\includegraphics[width=3cm]{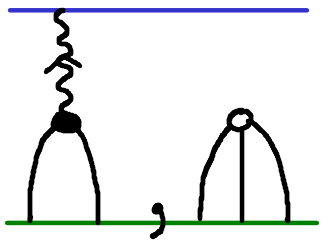}}}
\right]
&=&
3i\hbar
\stackrel{\hspace{-1.7cm}\Omega}{
\raisebox{-9mm}{\includegraphics[width=2.6cm]{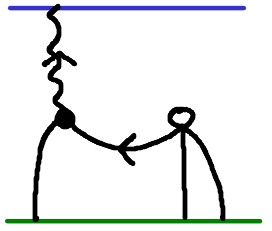}}}
+
3i\hbar
\stackrel{\hspace{1.5cm}\Omega}{
\raisebox{-9mm}{\includegraphics[width=2.2cm]{pope2}}}\\
&=&
6i\hbar
\stackrel{\hspace{1.5cm}\Omega}{
\raisebox{-9mm}{\includegraphics[width=2.2cm]{pope2}}}
-\: 6\hbar^2\stackrel{\hspace{-.13cm}\Omega}{
\raisebox{-9.7mm}
{\includegraphics[width=2cm]{PearlHarbor}}}
\, .
\end{eqnarray*}
Now note that
$$
\nabla_{(0)}=\:\stackrel{d\Pi}{\raisebox{-11mm}{\includegraphics[width=2.8cm]{d}}}
-\frac1{2i\hbar}
\stackrel{dT}{
\raisebox{-11.4mm}{\includegraphics[width=2.7cm]{ssddPi}}}\, ,$$
and
$$
\hat W_{(1)}=-\frac1{3!i\hbar}\,\raisebox{-6mm}{\includegraphics[width=1.4cm]{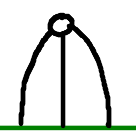}}\, ,\quad
\hat a_{(0)}=-\frac1{2i\hbar}
\stackrel{dX}{
\raisebox{-8mm}{\includegraphics[width=2.5cm]{ssddPi}}}\, ,\quad
\hat a_{(1)}=-\frac1{6i\hbar}\, 
\stackrel{\hspace{0cm}dT}{\raisebox{-9mm}{\includegraphics[width=2.6cm]{Pilat}}}\, ,
$$
where $dT^a:=(0,dt)=:d\Pi^a-dX^a$. Here $X^a=(\pi,\chi)$. Thus
$$
-
\big[\nabla_{(0)}
+\frac12\,  \hat a_{(0)}
,\hat W_{(1)}\big]
+\hat a_{(1)}=
\frac1{6i\hbar}
\stackrel{\hspace{0cm}dX}{
\raisebox{-9mm}{\includegraphics[width=2.6cm]{Pilat}}}
+\frac1{4i\hbar}
\stackrel{\hspace{1.5cm}dX}{
\raisebox{-8.8mm}{\includegraphics[width=2cm]{pope2}}}
+\frac14
\stackrel{\hspace{-.2cm}dX}{
\raisebox{-9mm}{\includegraphics[width=1.8cm]{PearlHarbor}}}\, .
$$
A check of this result is that it is antihermitean. This amounts to reversing the order of all the legs attached to the earth and then using the identity in Equation~\nn{racoon-drone} to restore these to the original pictures.

\medskip
Now, using the Poisson bracket identity in Equation~\nn{poissa} and the identity $dX^a = \partial_b \Pi^a e^b$, or in diagrams
$$
\stackrel{\hspace{-0cm}e}{
\raisebox{-6mm}{\includegraphics[width=2.3cm]{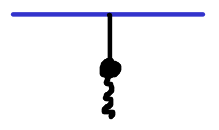}}}
=
\stackrel{\hspace{0cm}dX}{
\raisebox{-6.2mm}{\includegraphics[width=2.7cm]{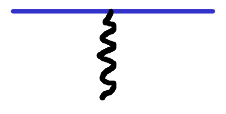}}}\, ,
$$
we have
$$
e^a \hat s_a = 
\stackrel{\hspace{-0cm}e}{
\raisebox{-9.6mm}{\includegraphics[width=1.9cm]{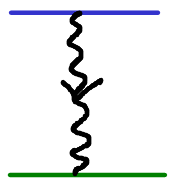}}}
=
\stackrel{\hspace{0cm}dX}{
\raisebox{-9.5mm}{\includegraphics[width=1.7cm]{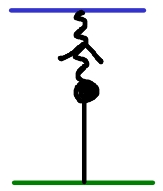}}}\, .
$$
Hence we compute
\begin{eqnarray*}
\left[
\stackrel{\hspace{-1.7cm}dX}{
\raisebox{-9mm}{\includegraphics[width=3cm]{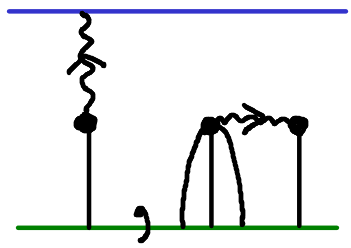}}}
\right]
&=&
-
i\hbar
\stackrel{\hspace{-0cm}dX}{
\raisebox{-9mm}{\includegraphics[width=2.6cm]{Pilat}}}
+3i\hbar
\stackrel{\hspace{-1.6cm}dX}{
\raisebox{-9mm}{\includegraphics[width=2.2cm]{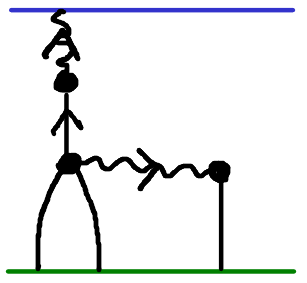}}}\\[2mm]
&=&
-4i\hbar
\stackrel{\hspace{-0cm}dX}{
\raisebox{-9mm}{\includegraphics[width=2.6cm]{Pilat}}}
-6i\hbar
\stackrel{\hspace{-1.3cm}dX}{
\raisebox{-9mm}{\includegraphics[width=2.4cm]{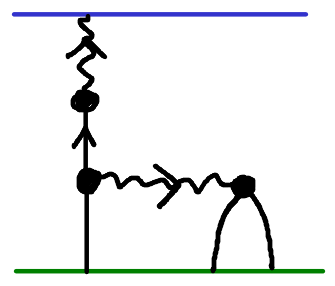}}}
+3(i\hbar)^2
\stackrel{\hspace{-.2cm}dX}{
\raisebox{-9mm}{\includegraphics[width=1.8cm]{PearlHarbor}}}\\[2mm]
&=&
-4i\hbar
\stackrel{\hspace{-0cm}dX}{
\raisebox{-9mm}{\includegraphics[width=2.6cm]{Pilat}}}
-6i\hbar
\stackrel{\hspace{1.5cm}dX}{
\raisebox{-9mm}{\includegraphics[width=2cm]{pope2}}}
-9(i\hbar)^2
\stackrel{\hspace{-.2cm}dX}{
\raisebox{-9mm}{\includegraphics[width=1.8cm]{PearlHarbor}}}
\end{eqnarray*}
Here we used Equations~\nn{identity} and~\nn{poissb} as well as the identity
$$
\stackrel{\hspace{-1.3cm}dX}{
\raisebox{-9mm}{\includegraphics[width=2.4cm]{Caligulina}}}
\:\:=\:\:
\stackrel{\hspace{1.5cm}dX}{
\raisebox{-9mm}{\includegraphics[width=2cm]{pope2}}}+
2i\hbar
\stackrel{\hspace{-.2cm}dX}{
\raisebox{-9mm}{\includegraphics[width=1.8cm]{PearlHarbor}}}
$$
\color{black}
Similarly
$$
\left[
\stackrel{\hspace{-1.7cm}dX}{
\raisebox{-9mm}{\includegraphics[width=3cm]{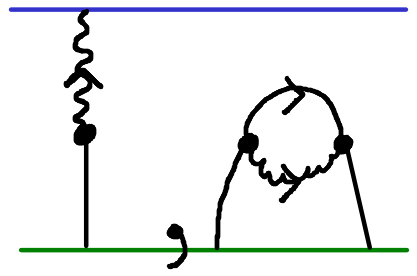}}}
\right]
\:\:=\:\:
-2i\hbar
\stackrel{\hspace{-.2cm}dX}{
\raisebox{-9mm}{\includegraphics[width=1.8cm]{PearlHarbor}}}\, .
$$
Thus we learn
$$
\operatorname{aHe}(\hat W_{(2)})=
-\frac1{24i\hbar}\, 
\raisebox{-2mm}{\includegraphics[width=2.4cm]{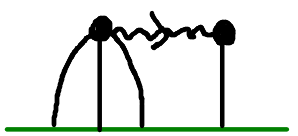}}
+\frac1{16}\, 
\raisebox{-4mm}{\includegraphics[width=1.4cm]{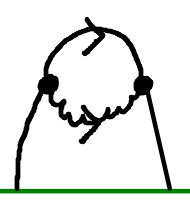}}
$$
Note that it is not difficult to see that the right hand side above  is antihermitean using the identity
$$
\raisebox{-4mm}{\includegraphics[width=1.3cm]{Bagpipes}}
:=-\raisebox{-4mm}{\includegraphics[width=2.2cm]{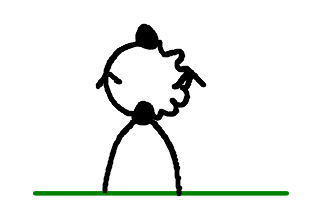}}\, .
$$
Let us define
$$
\raisebox{-6mm}{\includegraphics[width=1.6cm]{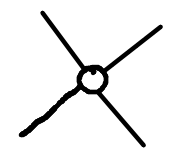}}
=\frac14
\left(
\raisebox{-4.5mm}{\includegraphics[width=2.1cm]{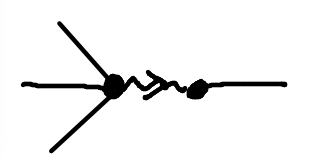}}
+
\raisebox{-5mm}{\includegraphics[width=1.9cm]{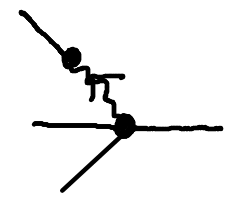}}
+
\raisebox{-9.5mm}{\includegraphics[width=1.9cm]{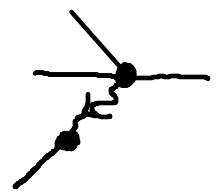}}
+\raisebox{-4.5mm}{\includegraphics[width=1.9cm]{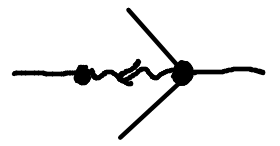}}
\right)\, ,
$$
where both sides of this diagrammatic picture define a totally symmetric rank four tensor.
Then is not difficult to check that
\begin{equation}\label{aHeW2}
\operatorname{aHe}(\hat W_{(2)})=
-\frac1{4!i\hbar}\, \raisebox{-4mm}{\includegraphics[width=2.6cm]{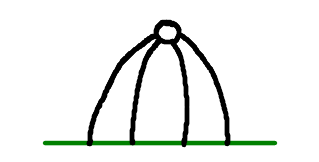}}\, .
\end{equation}

It is interesting to study even higher orders. For example, to    compute $\hat W_{(3)}$,  one can first consder the next order in Equation~\nn{gettinghe} which determines 
\begin{equation*}
\operatorname{He}(\hat W_{(3)})=
\frac12\Big(-
\hat W_{(1)}^3+
\hat W_{(1)}
\hat W_{(2)}
+
\hat W_{(2)}
\hat W_{(1)}
\Big)\, .
\end{equation*}
The remaining difficulty now is to calculate the heaven and earth diagrams for $\operatorname{aHe}(\hat W_{(3)})$.  Clearly this is possible but tedious. We have verified that in the model harmonic oscillator case where $V(q) =\frac12 q^2$, the obvious conjecture that symmetrizing the diagram
$$
\raisebox{-3.6mm}{\includegraphics[width=3.3cm]{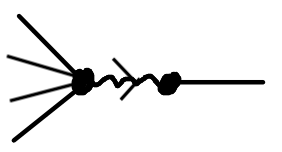}}
$$
computes $\operatorname{aHe}(\hat W_{(3)})$, in fact fails.
The problem of computing an all order expression for $\hat U$ is related to the problem of finding deformation quantizations for Poisson and symplectic structures. This problem is solved by Kontsevich's formality theorem~\cite{Kont}, which is neatly explained by the perturbative expansion of a  Poisson sigma model performed by  Cattaneo and Felder~\cite{Cat}. It is possible 
that those ideas could be applied to the computation of a formal asymptotic series result for the gauge transformation $\hat U$.

%
%
%
%

\subsection{Correlators}

We are now ready compute the evolution operator, which using Equation~\nn{woe} 
is given by 
$$
K(S_f,S_i;t_f-t_i):=
\langle S_f|\exp\big(-\tfrac {i(t_f-t_i)}\hbar\widehat H_q\big)
|S_i\rangle=\langle S_f|
P_\gamma \exp\Big(-\int_\gamma \widehat A\hh \Big)|S_i\rangle\, . 
$$
Here $\gamma$ is any path betwen $z_i=(0,q,t_i)$ and $z_f=(0,q,t_f)$ 
(in the $(p,q,t)$ coordinate system).
For the moment, let us focus on the quantity on the right hand side in the more general case that $z_i$ and $z_f$ are any two points in the manifold $Z$. In the previous section we computed the gauge transformation $\hat U$ such that
$$
\hat U
P_\gamma \exp\Big(-\int_\gamma \widehat A\hh \Big) \hat U^{-1}
=P_\gamma \exp\Big(-\int_\gamma \widehat A_{\rm D}\hh \Big)\, ,
$$
where $\widehat A_{\rm D}$ is the quantum connection potential in Darboux form as in Equation~\nn{AHOD}.
The above operator is easy to compute (see for example~\cite{Herla}): Consider $\gamma$ to be a path beginning at $z_i=(\uppi_i,\upchi_i,\uppsi_i)$ and ending at  $z_f=(\uppi_f,\upchi_f,\uppsi_f)$. Because the connection~$\nabla$ is flat, we may take $\gamma$ to particularly simple, for example one along which first only $\uppsi$ changes by amount $\uppsi_f-\uppsi_i$, then $\uppi$ by $\uppi_f-\uppi_i$ and finally $\upchi$ by $\upchi_f-\upchi_i$. This gives
$$
P_\gamma \exp\Big(-\int_\gamma \widehat A_{\rm D}\hh \Big)=\\
\exp\Big(\frac{\uppsi_f-\uppsi_i+\uppi_f(\upchi_f-\upchi_i)}{i\hbar}\Big)
\exp\Big((\upchi_f-\upchi_i)
\frac{\partial}{\partial S}\Big)
\circ\hh
\exp\Big(\frac{\uppi_f-\uppi_i}{i\hbar}\, S\Big)\, .
$$
Because  $\gamma$ is a path between $z_i=(0,q,t_i)$ and $z_f=(0,q,t_f)$, taking $H=\frac12 p^2+V(q)$ with $V(0)=0$, the above becomes (see Equation~\nn{covs}) simply a translation operator
 \begin{equation}\label{Ken}
P_\gamma \exp\Big(-\int_\gamma \widehat A_{\rm D}\hh \Big)=\exp\Big[{(t_f-t_i) \Big(\frac{V(q)}{i\hbar}-\frac{\partial}{\partial S}\Big)}\Big]=:\exp\Big(
-\tfrac {i(t_f-t_i)}\hbar\widehat H_{\rm D}\Big)\, .
\end{equation}
We also need to compute $\hat U |S_i\rangle$. Viewed as a wavefunction, the state $|S_i\rangle$
is represented by $\delta(S-S_i)$.
Now, remember that 
$$
\hat U = \hat U_0\big(1+\hat W_{(1)}+\hat W_{(2)}+\cdots\big) \, ,
$$
where $\hat U_0$ is the metaplectic representation of the matrix $U_0$ given in Equation~\nn{V}. Because $p$ and $q$ do not change along the path $\gamma$, at both the start and endpoint we have 
$$
U_0=\begin{pmatrix}
0&V'(q)\\-
\frac{1}{V'(q)}&0
\end{pmatrix}\, .
$$
Then using
Equation~\nn{letsgetmeta}, we have 
$$
\hat U_0^{-1}\circ
P_\gamma \exp\Big(-\int_\gamma \widehat A_{\rm D}\hh \Big)\circ
\hat U_0
=
\exp\Big[\frac{t_f-t_i}{i\hbar} \Big(V(q)
+ V'(q) S
\Big)\Big]\, .
$$
Next we imagine rewriting the remainder of the gauge transformation in normal order,{\it i.e.}
$$
1+\hat W_{(1)}+\hat W_{(2)}+\cdots=\colon {\mathcal W}_N(S,P)\colon\, , 
$$
where our normal ordering convention is 
$$
\colon S^k P^l\colon = S^k \big(\frac\hbar i \frac{\partial}{\partial S}\big)^l\, .
$$
Then, if $|P_i\rangle$ and $|P_f\rangle$ are eigenstates of the momentum operator $\frac\hbar i \frac{\partial}{\partial S}$, we have
\begin{multline*}
\langle P_f | 
P_\gamma \exp\Big(-\int_\gamma \widehat A\hh \Big)
|P_i\rangle=
e^{\frac{ V(q)\Delta t}{i\hbar}}
\int dS e^{-\frac{iV'(q) \Delta t\hh S}{\hbar}}\, 
{\mathcal W}_N^*(S,P_f) \hh
{\mathcal W}_N(S,P_i)\\[1mm]
=:
e^{\frac{ V(q)\Delta t}{i\hbar}}
{\mathcal F}\big[{\mathcal W}_N^*(S,P_f) \hh
{\mathcal W}_N(S,P_i)\big](-V'(q)\Delta t)\, , 
\end{multline*}
where $\Delta t:=t_f-t_i$ and ${\mathcal F}$ denotes the Fourier transform ${\mathcal F}\big[f(x)\big](k):=\int dx e^{\frac{ikx}\hbar} f(x)$.
To obtain the position space propagator two further Fourier transforms are needed, this yields
\begin{equation}\label{convolve}
K(S_i,S_f;\Delta t )=
{\mathcal F}\big[
{\mathcal W}_N^*(S,P') \hh
{\mathcal W}_N(S,P)
\big](-S_f,S_i,-V'(q)\Delta t)\, ,
\end{equation}
where the variables $(P',P,S)$ are the respective Fourier dual variables to $(S_f,\!S_i,\!\!-V'(q)\Delta t)$.
The above Fourier convolution result for quantum mechanical propagators is a very strong result, but it comes with a caveat which we now describe.

The equality in the  above convolution result assumes that we can find an exact expression for the operator $\hat U_1$.
However, in general, only an asymptotic series expression for $\hat U_1$ will exist. 
To see why this is, it is useful to study the harmonic oscillator example~$V(q)= \frac 12 q^2$. In that case the operator $\hat U_1$ must solve the condition
$$
\hat U_1^{-1}\circ
\exp\Big[\frac{\Delta t}{i\hbar} \Big(\frac12 q^2 + q S\Big)\Big]\circ \hat U_1 = 
\exp\Big[\frac{\Delta t}{i\hbar} \Big(-\frac{\hbar^2}2 
\frac{\partial^2}{\partial S^2} + \frac12 (q+S)^2
\Big)\Big]\, .
$$
Because the spectrum of the harmonic oscillator Hamiltonian is discrete, while that of the operator $q S$ is continuous, no unitary operator $\hat U_1$ 
solving the above operator equation can exist. However, there is a asymptotic series solution.
Let us complete our study of the quantum Darboux theorem by demonstratng how these asymptotics work for the harmonic oscillator. 

The first three asymptotic orders of the operator $\hat U$ were computed for general models using  heaven and earth diagrams in Section~\ref{H+E}. To explicate this series it is useful to introduce the new variable
$$
\sigma:=\frac{S}{\sqrt \hbar}\, ,\qquad 
\varepsilon:=\frac{\sqrt \hbar}q\, ,
$$
in terms of which the display before last becomes
\begin{equation}\label{inter}
\hat U_1^{-1}\circ
\exp\Big[-i\Delta t \Big(\frac1{2 \varepsilon^2} + \frac{\sigma}{\varepsilon}\Big)\Big]\circ \hat U_1 = 
\exp\Big[-i\Delta t\Big(
\frac1{2 \varepsilon^2} 
 + \frac{\sigma}{\varepsilon} -\frac12 
\frac{\partial^2}{\partial \sigma^2} + \frac12 \sigma^2\Big)\Big]\, .
\end{equation}
For the harmonic oscillator, the pair
$
\Pi= (\pi, \phi)=(\frac12 p^2 + \frac12 q^2,-\arctan \frac pq)
$. Hence, at the start point of a path $\gamma$ with $p=0$, the operator $\hat W_{(1)}$ given in Equation~\nn{What1}
becomes
$$
\hat W_{(1)}=
\frac{\varepsilon}{3!}
\Big(
\frac{\partial^3}{\partial \sigma^3}
-3 \sigma^2
\frac{\partial}{\partial \sigma}
+3\sigma\Big)
\, ,
$$
while the anti-Hermitean part of $\hat W_{(2)}$ given in Equation~\nn{aHeW2} is
$$
\operatorname{aHe}(\hat W_{(2)})=
\frac{\varepsilon^2}{4!}\Big(
2\sigma \frac{\partial^3}{\partial \sigma^3}
+6 \sigma^3 
\frac{\partial}{\partial \sigma}
+3 \frac{\partial^2}{\partial \sigma^2}
+9 \sigma^2
\Big)\, .
$$
Thus
\begin{multline}\label{gauge}
\hat U_1=1
+\hat W_{(1)}
+\operatorname{aHe}(\hat W_{(2)})
+
\frac12
\hat W_{(1)}^2
+
{\mathcal O}(\varepsilon^3)\\[2mm]
=
1+
\frac{\varepsilon}{3!}
\Big(
\frac{\partial^3}{\partial \sigma^3}
-3 \sigma^2
\frac{\partial}{\partial \sigma}
+3\sigma\Big)\hspace{6.4cm}
\\[1mm]
 + \frac{\varepsilon^2}{4!} \Big(
\frac13\frac{\partial^6}{\partial\sigma^6}
-2\sigma^2 
\frac{\partial^4}{\partial\sigma^4}
+3 \sigma^4 
\frac{\partial^2}{\partial\sigma^2}
-6 \sigma
\frac{\partial^3}{\partial\sigma^3}
+18 \sigma^3
\frac{\partial}{\partial\sigma}
-6\frac{\partial^2}{\partial\sigma^2}
             + 15\sigma^2                    \Big)\hspace{-1cm}
             \\[2mm]
+{\mathcal O}(\varepsilon^3)\, .\hspace{10.6cm}
\end{multline}
It is not difficult to check that
$$
\hat U^{\dagger}_1 \hat U_1 = 1 +{\mathcal O}(\varepsilon^3)\, ,
$$
and that
\begin{equation}\label{conjugate}
\hat U^{\dagger}_1 
\circ
\sigma
\circ
\hat U_1
=
\sigma 
+\frac\varepsilon2
\Big(
 -
\frac{\partial^2}{\partial \sigma^2} + \sigma^2 \Big)
+
{\mathcal O}( \varepsilon^3)\, .
\end{equation}
Note that as mentioned in the previous section, it can be easily checked that the antiHermitean part of $\hat W_{(3)}$ is not given solely by the five point diagram displayed there.

Now for sake of generality, imagine  that we had solved for $\hat U_1$ asymptotically to order~$\varepsilon^k$ in the above display. Then, introducing the variable
$$
\delta:=\frac{\Delta t}\varepsilon\, ,
$$
we can develop an asymptotic series expansion in $\varepsilon$
in the scaling limit where $\delta$ is held fixed for the evolution operator on the left hand side of Equation~\nn{inter}
$$
e^{-\frac i2 \frac \delta\varepsilon}\:
\hat U_1^{-1}\circ
\exp\Big[-i \delta \sigma\Big]\circ \hat U_1 = 
e^{-\frac i2 \frac \delta\varepsilon}\: \exp\Big[-i\delta\Big\{\sigma 
+\frac\varepsilon2
\Big(
 -
\frac{\partial^2}{\partial \sigma^2} + \sigma^2 \Big)\Big\}\Big]
+{\mathcal O}(\varepsilon^k)\, .
$$
Equation~\nn{conjugate} ensures that the gauge transformation given in~\nn{gauge} solves the above displayed equality of asymptotic series for the case $k=3$.

\section*{Acknowledgements}
We thank Gabriel Herczeg for a collaboration in the early stages of this work. We are also very appreciative of extensive discussions with Roger Casals.
A.W. was supported in part by  
Simons Foundation Collaboration Grants for Mathematicians ID 317562 and 686131-0.


\begin{thebibliography}{99}


\bibitem{Dupre} M.J. Dupr\'e,
J. Functional Analysis {\bf 15}  244
(1974).


 \bibitem{Krysl}
 S. Kr\'ysl,  
 Diff. Geom. Appl. {\bf 33} 290
 (2014)
 (arXiv:1304.5704 [math.DG]).  

\bibitem{Her}
G. Herczeg and A. Waldron, 
Phys. Lett. B{\bf 781}, 
312
 (2018),
 arXiv:1709.04557.
 
 \bibitem{Catt}
 A. S. Cattaneo, P. Mnev and K. Wernli, ``Constrained systems, generalized Hamilton--Jacobi actions, and quantization'', arXiv:2012.13720.

\bibitem{Herla}
G. Herczeg, E. Latini and A. Waldron
Arch. Math. (Brno) {\bf 54} no. 5, 
281
(2018),  
arXiv:1805.11731.







\bibitem{He}
Z. He, ``Odd Dimensional Symplectic Manifolds'', MIT Ph.D. Thesis, 2010.

\bibitem{Geiges}
H. Geiges ``An Introduction to Contact Topology'',
Cambridge University Press 2008; 
P.~{\v{S}}evera, 
J. of
  Geometry and Physics {\bf 29} 235
  (1999);
  S.~G. Rajeev, 
Annals of Physics {\bf 323} 768
(2008).

\bibitem{Fed}
B.V. Fedosov, 
J. Differential Geometry {\bf 40} 213
(1994).  


\bibitem{Grig}
M.A. Grigoriev and S.L. Lyakhovich
Commun. Math.Phys.  {\bf 218} 437 (2001) (hep-th/0003114). See also G. Barnich and M. Grigoriev, 
A. Semikhatov and I. Tipunin,
 Commun. Math. Phys. {\bf 260} 147
 (2005)  (hep-th/0406192).

\bibitem{BFVsecond}
I. Batalin, E. Fradkin, and T. Fradkina, 
Nucl. Phys. B{\bf 314} 
158
(1989);
I. A. Batalin and I. V. Tyutin, 
Int. J. of Mod. Phys. A{\bf 6} 3255
(1991);
%
%
I.~Batalin, E.~Fradkin, and T.~Fradkina, 
  Nucl. Phys. B{\bf 314}158
  (1989).

\bibitem{Witten}
E. Witten
J. Diff. Geom. {\bf 17},
 661
 (1982).

\bibitem{Kostant}
B. Kostant, 
``Symplectic spinors'' in Symposia Mathematica, Vol. XIV (Convegno di Geometria Simplettica e Fisica Matematica, INDAM, Rome, 1973), 139--152, Academic Press, London, 1974.

\bibitem{Fitz}
S.~Fitzpatrick, 
  J. Geometry and Physics
  {\bf 61} 2384
 (2011).


\bibitem{Kont}
M. Kontsevich, 
Lett. Math. Phys. {\bf 66}, 
157
(2003), 
arXiv:q-alg/9709040; 
Lett. Math. Phys. {\bf 48}, 
35
(1999),
arXiv:math/9904055 .

\bibitem{Cat}
A. Cattaneo and  G. Felder, 
Commun. Math. Phys. {\bf 212}, 591
(2000), arXiv:math.QA/9902090;
Mod. Phys. Lett. A{\bf 16},
179
(2001), arXiv:hep-th/0102208.
















\bibitem{BFV}
E.~S. Fradkin and G.~Vilkovisky, 
Phys. Lett. B{\bf 55} 224
(1975);
I.~A. Batalin and G.~A. Vilkovisky, 
Phys. Lett. B{\bf 69} 309
(1977);
E.~S. Fradkin and T.~Fradkina, 
Phys. Lett. B{\bf 72} 343
(1978);
I.~Batalin and E.~S. Fradkin,  
La Rivista del
  Nuovo Cimento {\bf 9} 1 
  (1986).

%
%
%
%
%
%
%
%
%
%
%
%
%
%
%
%
%
%
%
%
%
%
%
%
%
%
%
%
%
%
%
%
%
%
%
%
%
%
%
%
%
%
%



\end{thebibliography}
\end{document}